\documentclass[preprint,5p]{elsarticle}

\usepackage{graphicx,amssymb}
\usepackage{epsf,psfig}

\journal{New Astronomy}

\begin{document}

\begin{frontmatter}

\title{A new dynamical model for the study of galactic structure}

\author{Euaggelos E. Zotos\corref{}}

\address{Department of Physics, \\
Section of Astrophysics, Astronomy and Mechanics, \\
Aristotle University of Thessaloniki \\
GR-541 24, Thessaloniki, Greece}

\cortext[]{Corresponding author: \\
\textit{E-mail address}: evzotos@astro.auth.gr (Euaggelos E. Zotos)}

\begin{abstract}

In the present article, we present a new gravitational galactic model, describing motion in elliptical as well as in disk galaxies, by suitably choosing the dynamical parameters. Moreover, a new dynamical parameter, the $S(g)$ spectrum, is introduced and used, in order to detect islandic motion of resonant orbits and the evolution of the sticky regions. We investigate the regular or chaotic character of motion, with emphasis in the different dynamical models and make an extensive study of the sticky regions of the system. We use the classical method of the Poincar\'{e} $(r-p_r)$ phase plane and the new dynamical parameter of the $S(g)$ spectrum. The L.C.E is used, in order to make an estimation of the degree of chaos in our galactic model. In both cases, the numerical calculations, suggest that our new model, displays a wide variety of families of regular orbits, compared to other galactic models. In addition to the regular motion, this new model displays also chaotic regions. Furthermore, the extent of the chaotic regions increases, as the value of the flatness parameter $b$ of the model increases. Moreover, our simulations indicate, that the degree of chaos in elliptical galaxies, is much smaller than that in dense disk galaxies. In both cases numerical calculations show, that the degree of chaos increases linearly, as the flatness parameter $b$ increases. In addition, a linear relationship between the critical value of angular momentum and the $b$ parameter if found, in both cases (elliptical and disk galaxies). Some theoretical arguments to support the numerical outcomes are presented. Comparison with earlier work is also made.

\end{abstract}


\end{frontmatter}

\section{Introduction}

It is very interesting, that scientists interested in Dynamical Astronomy, are often model constructors. Dynamical models, are very useful tools for the study of the behavior of orbits in a stellar dynamical system, such as a galaxy. A dynamical model is usually a mathematical expression, giving the potential as a function of the distance from center of the galaxy. Although, sometimes instead of the potential, the density or the gravitational force as a function of the distance, is given.

Important and also interesting models, have been built up during the last three decades, when the idea that the elliptical galaxies were considered as rotationally flattened systems, has been abandoned. Two main categories of models, were used for the study of the galactic motion. (1) The local models, built up of harmonic oscillators (see Saito and Ichimura 1974, Caranicolas 1984, 1993, 1994, Deprit 1991, Elipe et al. 1995, Elipe 1999, Karanis and Caranicolas 2002) and (2) The global mass models, describing global motion (see Clutton-Brock et al. 1976, Martinet and Pfenniger 1986, Carlberg and Innanen 1987, Patsis and Zachilas 1990, Flynn et al. 1996, Fellhauer et al. 2006). A detailed description for both the above two kinds of galactic dynamical models can been found, by the reader in Binney and Tremaine (2008).

In several papers, we see that low angular momentum stars in galaxies with massive nuclei, are scattered to much higher scale heights, displaying chaotic motion (see Caranicolas \& Innanen 1991, Caranicolas 1997, Karanis \& Caranicolas 2001, Caranicolas \& Papadopoulos 2003, Caranicolas \& Zotos 2010). All the above research, not only describes the chaotic character of orbits of low angular momentum stars, but also gives the main reason responsible for this behavior. The reason is the presence of the strong   force, which comes from the massive nucleus. The dynamical models, used in all previous papers were global models. In Caranicolas \& Innanen (1991) and Caranicolas (1997), they have used the mass model introduced by Carlberg \& Innanen (1987), in Karanis \& Caranicolas (2001), they have used a logarithmic potential, while in Caranicolas \& Papadopoulos (2003) a time - dependent model was used.

The most characteristic result, derived in all cases, was the linear dependence of the critical value of angular momentum and the mass of the nucleus. From our point of view, the background of this series of paper, in combination of other orbit calculations, was that managed to find and present, both numerically and semi - theoretically, relationships connecting the physical parameters of the dynamical system, such as the angular momentum and the degree of chaos in elliptical and disk galaxies.
\begin{figure*}[!tH]
\centering
\resizebox{0.9\hsize}{!}{\rotatebox{0}{\includegraphics*{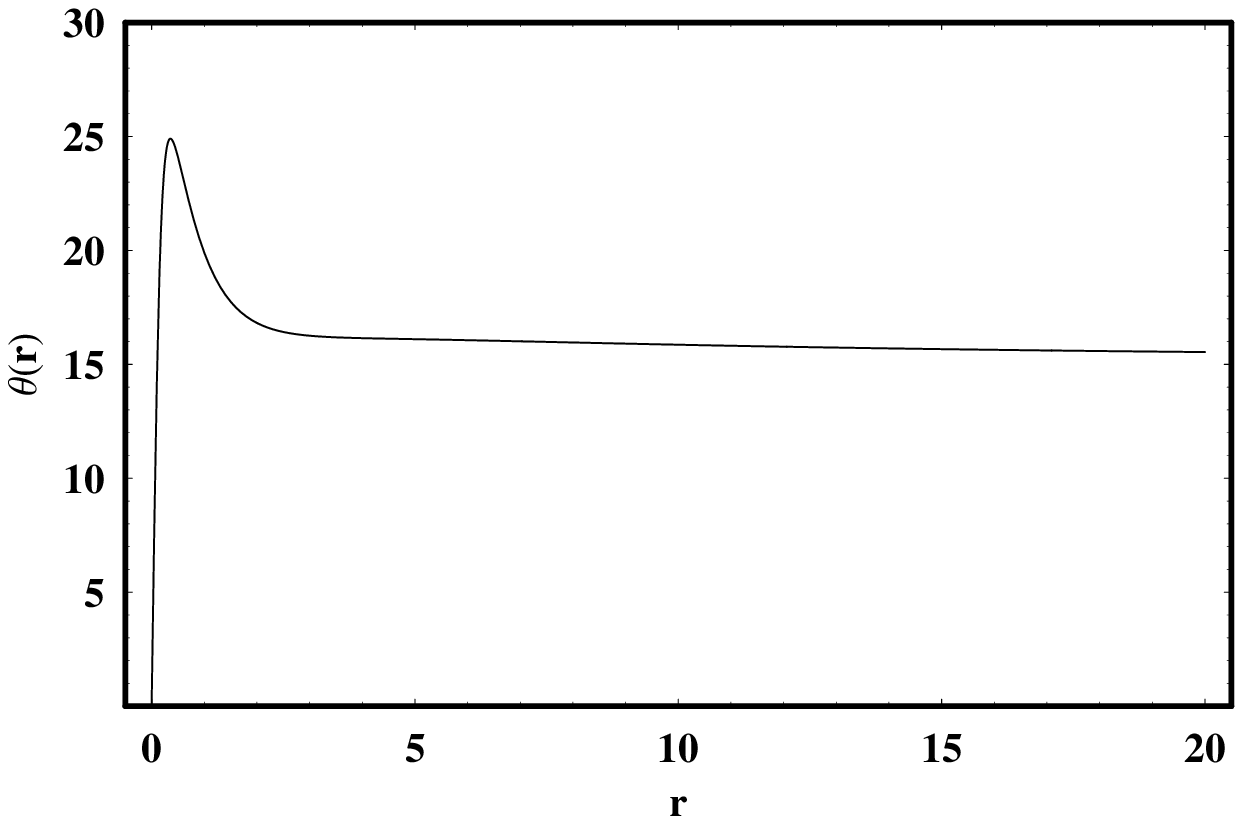}}\hspace{1cm}
                         \rotatebox{0}{\includegraphics*{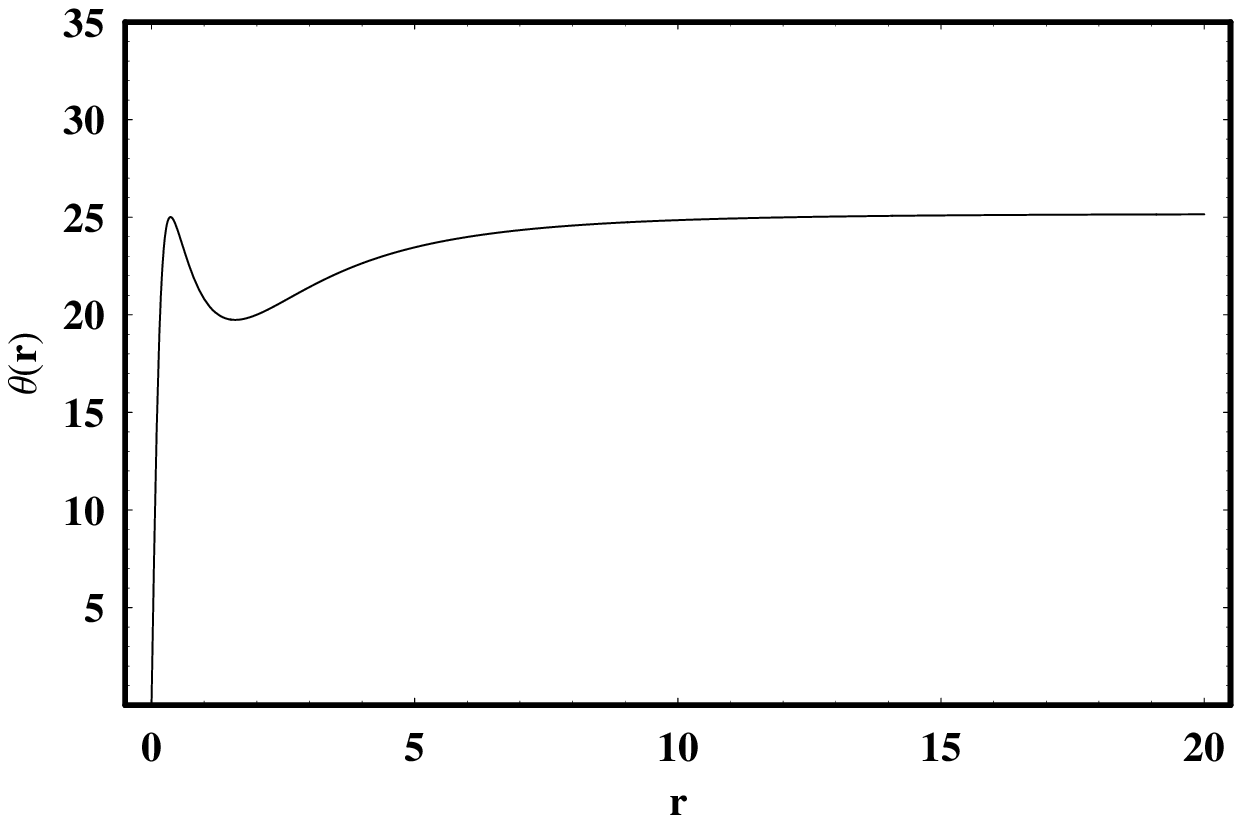}}}
\vskip 0.1cm
\caption{(a-b): (a-left): Rotation curve for the elliptical galaxy and (b-right): Rotation curve for the disk galaxy. The values of all parameters are given in text while, $\Theta$ is the circular velocity in the galactic plane $(z=0)$.}
\end{figure*}
\begin{figure*}[!tH]
\centering
\resizebox{0.9\hsize}{!}{\rotatebox{0}{\includegraphics*{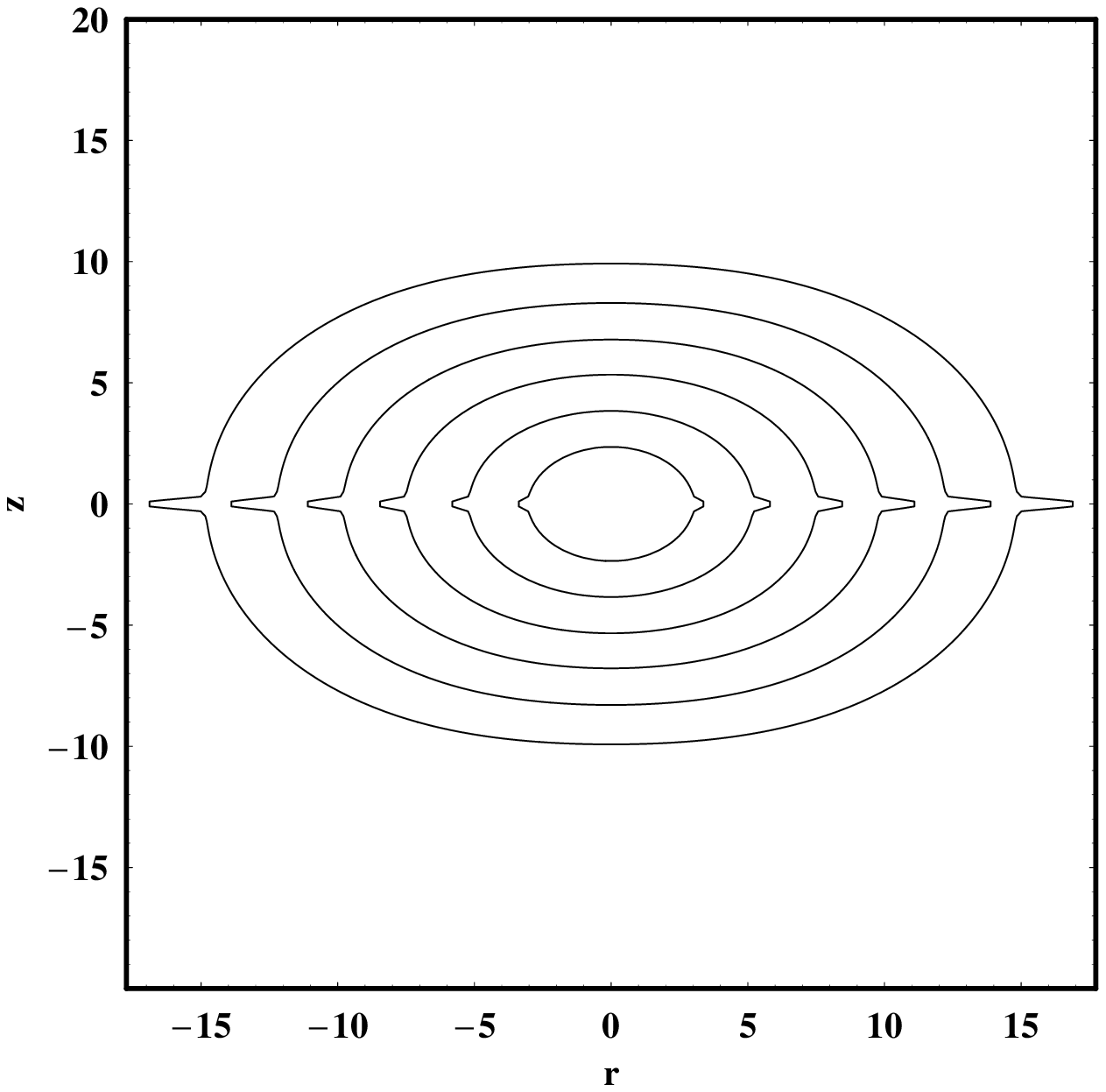}}\hspace{1cm}
                        \rotatebox{0}{\includegraphics*{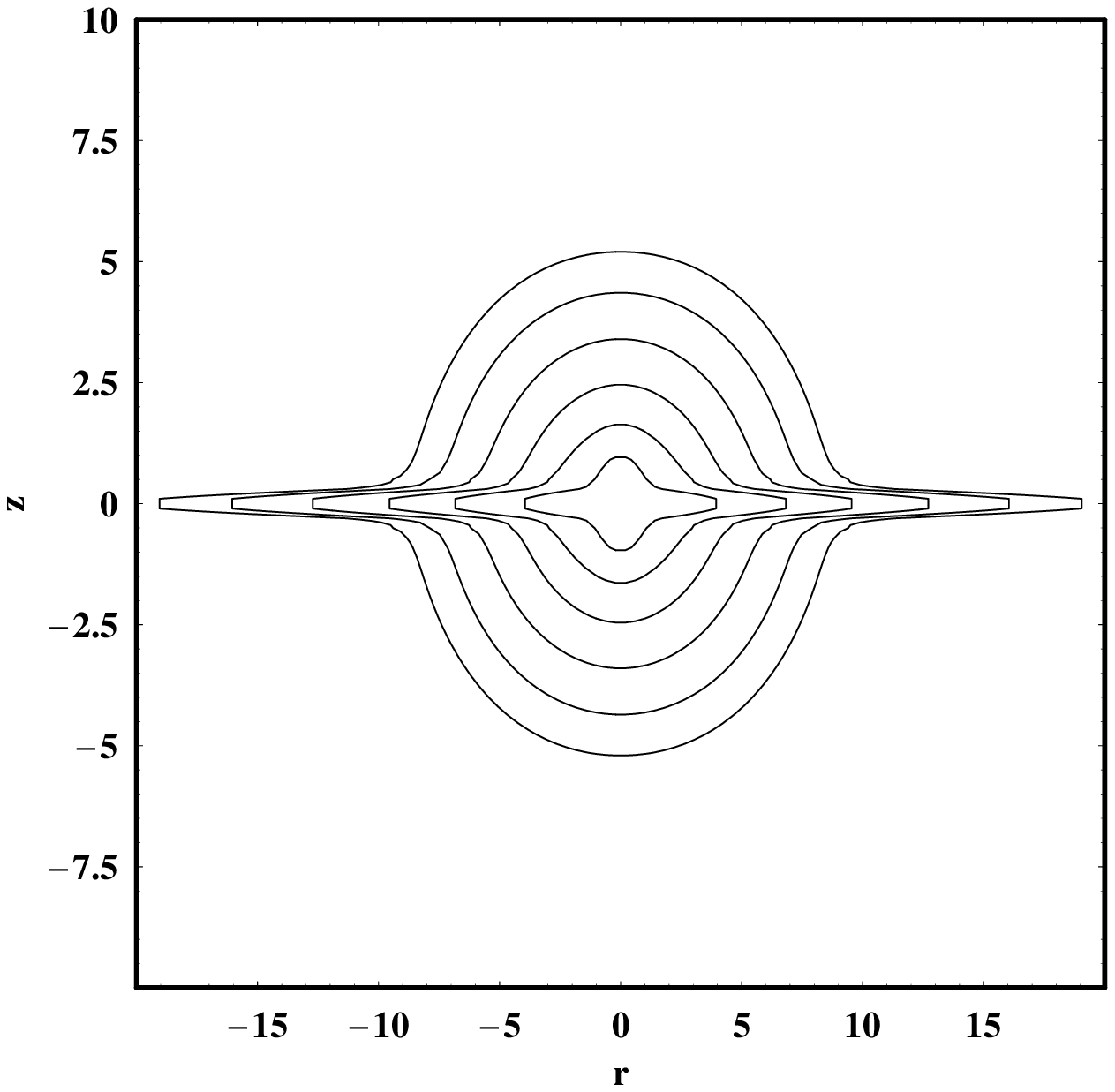}}}
\vskip 0.1cm
\caption{(a-b): (a-left): Density contours for the elliptical galaxy and (b-right):
Density contours for the disk galaxy. The values of all parameters are given in text.}
\end{figure*}
\begin{figure*}[!tH]
\centering
\resizebox{0.9\hsize}{!}{\rotatebox{0}{\includegraphics*{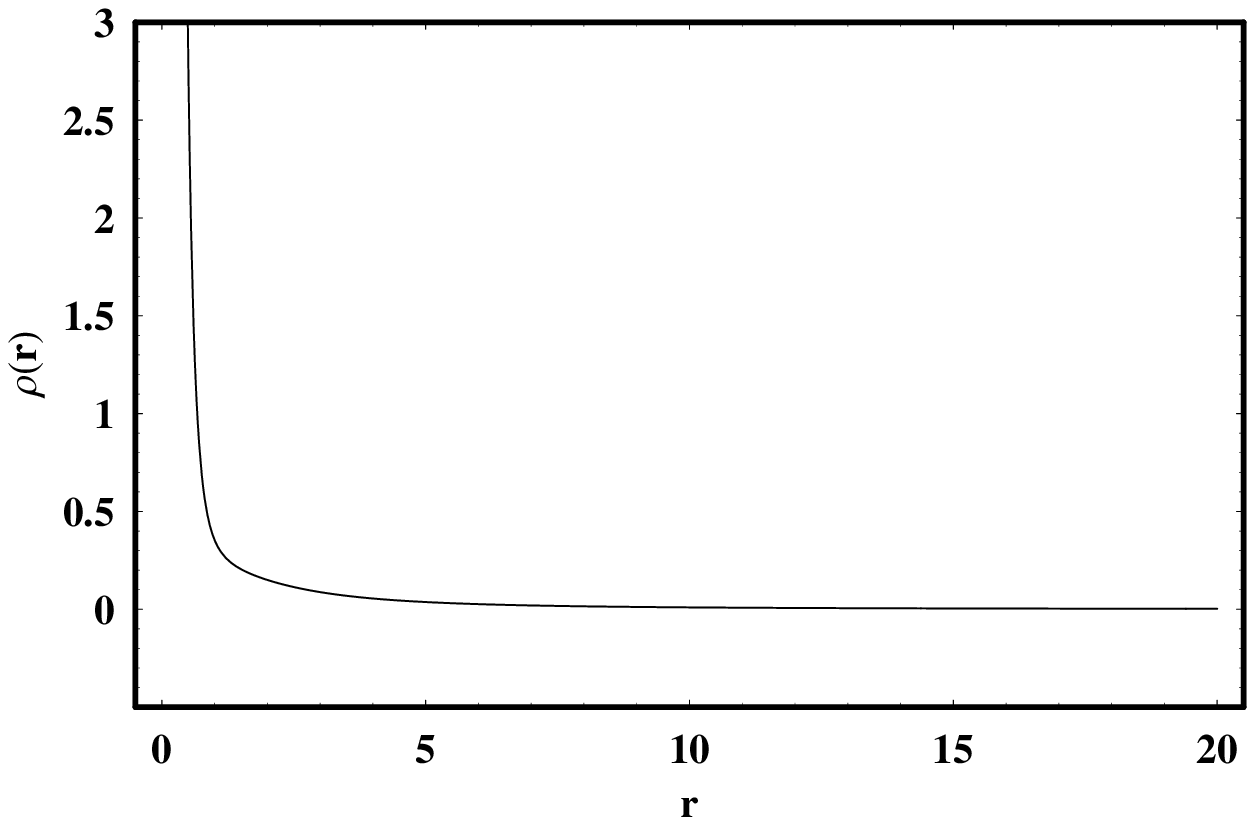}}\hspace{1cm}
                         \rotatebox{0}{\includegraphics*{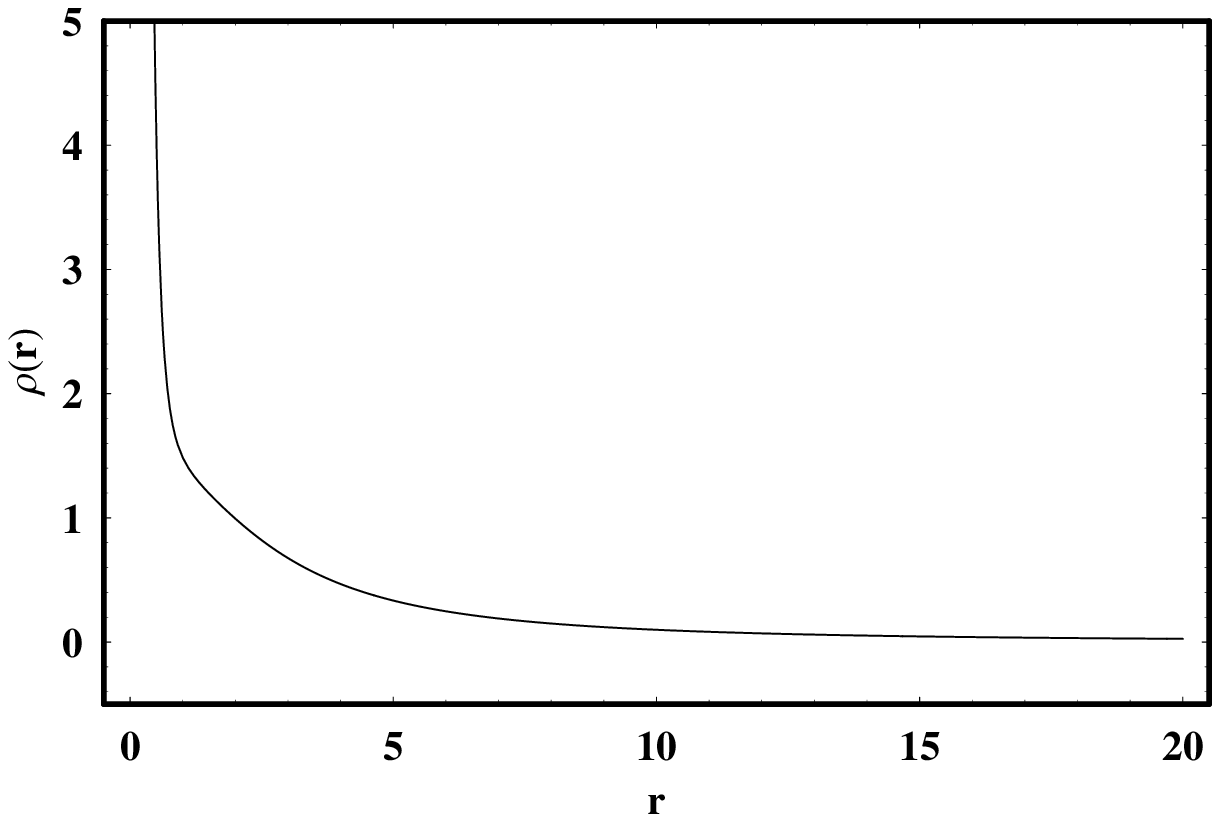}}}
\vskip 0.1cm
\caption{(a-b): The mass density in the galactic plane $z=0$, (a-left): for the elliptical galaxy and (b-right) for the disk galaxy.}
\end{figure*}

The paper is organized as follows: In Section 2, we present our new dynamical model, which is a combination of the logarithmic model and the Miyamoto - Nagai model (1975). In  Section 3, we study the character of orbits in two different cases (i) Our model describes an elliptical galaxy and (ii) our model describes a flat disk galaxy. Moreover, a comparison between those two cases is investigated. In Section 4, we introduce and use the new $S(g)$ spectrum. In the next Section 5, we present some semi-theoretical arguments, in order to support our numerically obtained outcomes. Finally, in the last Section a discussion and the conclusions of this research are presented.

\section{Presentation of the dynamical model}

In the present paper, we shall study the properties of motion in a galaxy described, by the potential
\begin{equation}
V_g\left(r,z\right)=\frac{\upsilon_0^2}{2} \ln \left[r^2+\left(a+\sqrt{h^2+bz^2}\right)^2+c^2\right],
\end{equation}
were $r, z$ are the cylindrical coordinates, $c$ is the scale length of the galaxy, while $a$ and $h$ are parameters regarding the structure of the dynamical system. The parameter $\upsilon_0$, is used for the consistency of the galactic units. The parameter $b$ is very important, because it represents the flatness of the galaxy. The shape of galaxy is spherical or oblate when $1 \leq b < 2$ or prolate when $0.1 \leq b < 1$. In the present paper, we shall study in detail, the case of two oblate galactic systems (disk and elliptical galaxy) with the same value of flatness parameter $(b=1.3)$. Moreover, we shall compare the relationships between other parameters of the dynamical systems and the flatness parameter, when $1 < b < 2$. There are three main line of arguments, justify our choice: (i). It is a simple dynamical model, describing motion in a galaxy with an axis and a plane of symmetry. (ii). This potential can describe motion, in multiple models, from an elliptical galaxy $\left(a\ll b \right)$, to a flat disk system. (iii). Combining potential (1) with the potential of a spherically symmetric nucleus, we can obtain a new model, displaying a wide variety of regular orbits, together with chaotic regions.

Adding to potential (1) a spherically symmetric nucleus, we obtain the total potential
\begin{equation}
V_{tot}\left(r,z\right)=V_g\left(r,z\right)+V_n\left(r,z\right),
\end{equation}
where
\begin{equation}
V_n\left(r,z\right)=\frac{-M_n}{\sqrt{r^2+z^2+c_n^2}},
\end{equation}
while, $M_n$ is the mass and $c_n$ represents the scale length of the nucleus. The Plummer sphere we use in order to increase the central mass, has been applied many times in the past, in order to study the effect of the introduction of a central mass component in a galaxy (see Hasan and Norman 1990, Hasan et al. 1993). In this work, we use a system of galactic units, where the unit of mass is $2.325 \times 10^7 M_\odot$, the unit of length is $1kpc$ and the unit of time is $0.997748 \times 10^8 yr$. The velocity unit and the energy unit (per unit mass) are $10 km/s$ and $100 (km/s)^2$ respectively, while $G$ is equal to unity. In these units, we use the values: $c=2.5, M_n=400, c_n=0.25$, while $\upsilon_0, a, h, b$ are treated as parameters.

Figure 1a, shows the rotation curve of the galaxy described by potential (2), when: $\upsilon_0=15, a=0.1, b=1.3, h=0.12$. In this case, we have an elliptical galaxy, with a small circular velocity, except in the central region and a low mass density distribution. At $r=8.5, z=0$, we find a circular velocity $\Theta=160 km/s$ and a mass density $\rho=0.014 M_\odot /pc^3$. On the contrary, when $\upsilon_0=25, a=1.2, b=1.3, h=0.2$, equation (2) describes a flat disk system. The corresponding rotation curve is shown in Fig. 1b. Here things are very different, because at $r=8.5, z=0$, we obtain a circular velocity $\Theta=246 km/s$ and a mass density $\rho=0.134 M_\odot /pc^3$. It is very useful, to compute the galactic density $\rho \left(r,z\right)$, derived from the total potential (2) using the Poisson's equation
\begin{equation}
\rho \left(r,z\right)=\frac{1}{4\pi G}\nabla ^2 V_{tot}\left(r,z\right).
\end{equation}

Figures 2a-b, shows the contours $\rho \left(r,z\right)=const$, for the elliptical and disk galaxy respectively. For the elliptical galaxy in Fig. 2a, the contours are: (0.057, 0.021, 0.01, 0.0058, 0.0037, 0.0025), while for the disk model in Fig.2b the contours are: (0.34, 0.14, 0.076, 0.044, 0.028, 0.02). In Figures 3a-b, we can observe the plots of the mass density in the galactic plane $\rho \left(r,z=0\right)$, as a function of the radius $r$, for the elliptical and disk galaxy respectively. Due to the fact that, the potential is logarithmic, for large values of $r$ and $z$, the mass density varies like $1/r^2$ and $1/z^2$ respectively. This means that the total mass $M(R)$, enclosed in a sphere of radius $R$ increases linearly with distance. This explains why circular velocity profiles, shown in Fig. 1a-b are flat. Here we must point out, that our gravitational potential is truncated at $R_{max}=20 kpc$, otherwise the total mass of the galaxies modeled by this potential are infinite, which is obviously not physical.

\section{Structure of the phase plane - Orbit calculations}

As the total potential $V_{tot}\left(r,z\right)$ is axially symmetric and the $L_z$ component of the angular momentum is conserved, the dynamical structure of the system, can be studied using the effective potential
\begin{equation}
V_{eff}\left(r,z\right)=\frac{L_z^2}{2r^2}+V_{tot}\left(r,z\right),
\end{equation}
in order to study the motion in the $(r,z)$ plane. The equations of motion are
\begin{eqnarray}
\dot{r}=p_r, \ \ \ \dot{z}=p_z, \nonumber \\
\dot{p_r}=-\frac{\partial \ V_{eff}}{\partial r}, \ \ \
\dot{p_z}=-\frac{\partial \ V_{eff}}{\partial z},
\end{eqnarray}
where the dot indicates derivative with respect to time.

The corresponding Hamiltonian can be written as
\begin{equation}
H=\frac{1}{2} \left( p_r^2+p_z^2 \right) + V_{eff}\left(r,z\right)=E,
\end{equation}
where $p_r$ and $p_z$, are the momenta per unit mass, conjugate to $r$ and $z$ respectively, while $E$ is the numerical value of the Hamiltonian, which is conserved. Equation (7) is an integral of motion, which indicates that the total energy of the test particle is conserved. The Hamiltonian (7), also describes the motion in the $(r,z)$ meridian plane, rotating at the angular velocity
\begin{equation}
\dot{\phi}=\omega =\frac{L_z}{r^2}.
\end{equation}

In this Section, we shall use the classical method of the Poincar\'{e} $(r,p_r)$, $z=0, p_z>0$ phase plane, in order to determine the regular or chaotic nature of motion. If we set $z=p_z=0$ in equation (7), we obtain the limiting curve in the $(r-p_r)$ phase plane, which is the curve containing all the invariant curves, for a given value of the energy integral $E$. The limiting curve of the dynamical system corresponds to
\begin{equation}
\frac{1}{2}p_r^2+V_{eff}(r)=E.
\end{equation}

Orbit calculations, are mainly based on the numerical integration of the equations of motion (6), which was made using a Bulirsh - Stoer numerical integration Fortran code, with double precision subroutines. The accuracy of the calculations, was always checked by the constancy of the energy integral, which was conserved up to the twelfth significant figure.

Two cases will be studied in this investigation: (a) The model (1) represents an \textbf{elliptical galaxy} and (b) The model (1) represents a \textbf{flat disk galaxy}.

\subsection{Elliptical galaxy model}

Figure 4 shows the structure of the Poincar\'{e} phase plane, when $\upsilon_0=15, a=0.1, b=1.3, h=0.12$, for the Hamiltonian (7). It is evident that, for the above values of the dynamical parameters, we study the orbital structure of an elliptical galaxy. The value of the energy is $E=486$ and the value of the angular momentum is $L_z=10$.
\begin{figure}[!tH]
\resizebox{\hsize}{!}{\rotatebox{270}{\includegraphics*{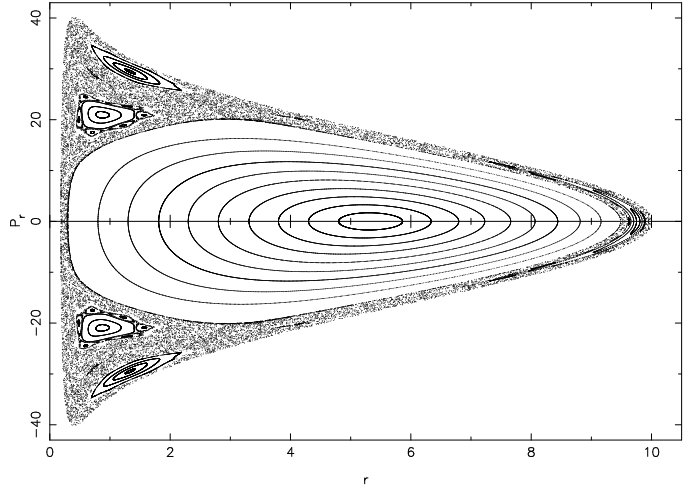}}}
\caption{The $(r-p_r)$ phase plane for the elliptical galaxy, when $E=486$. See text for details.}
\end{figure}
\begin{figure*}[!tH]
\centering
\resizebox{0.8\hsize}{!}{\rotatebox{0}{\includegraphics*{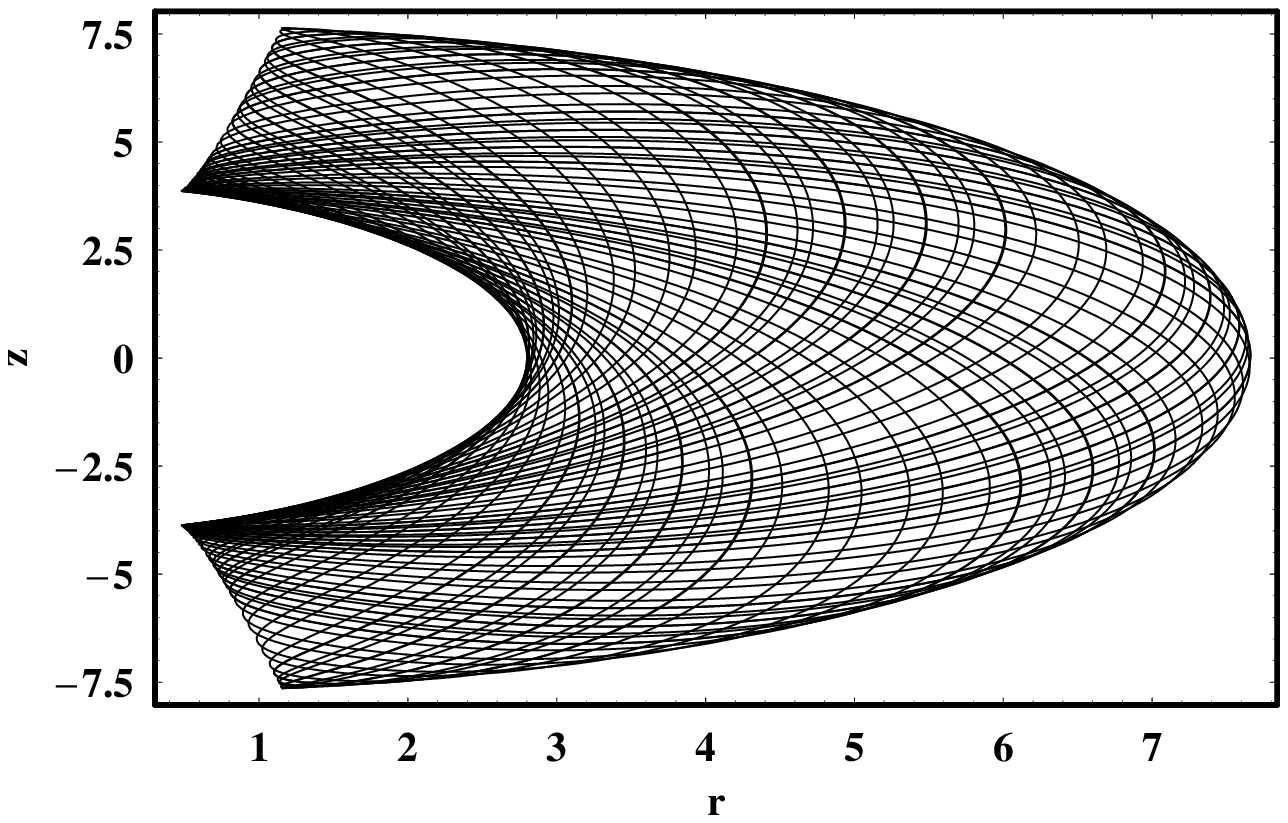}}\hspace{1cm}
                         \rotatebox{0}{\includegraphics*{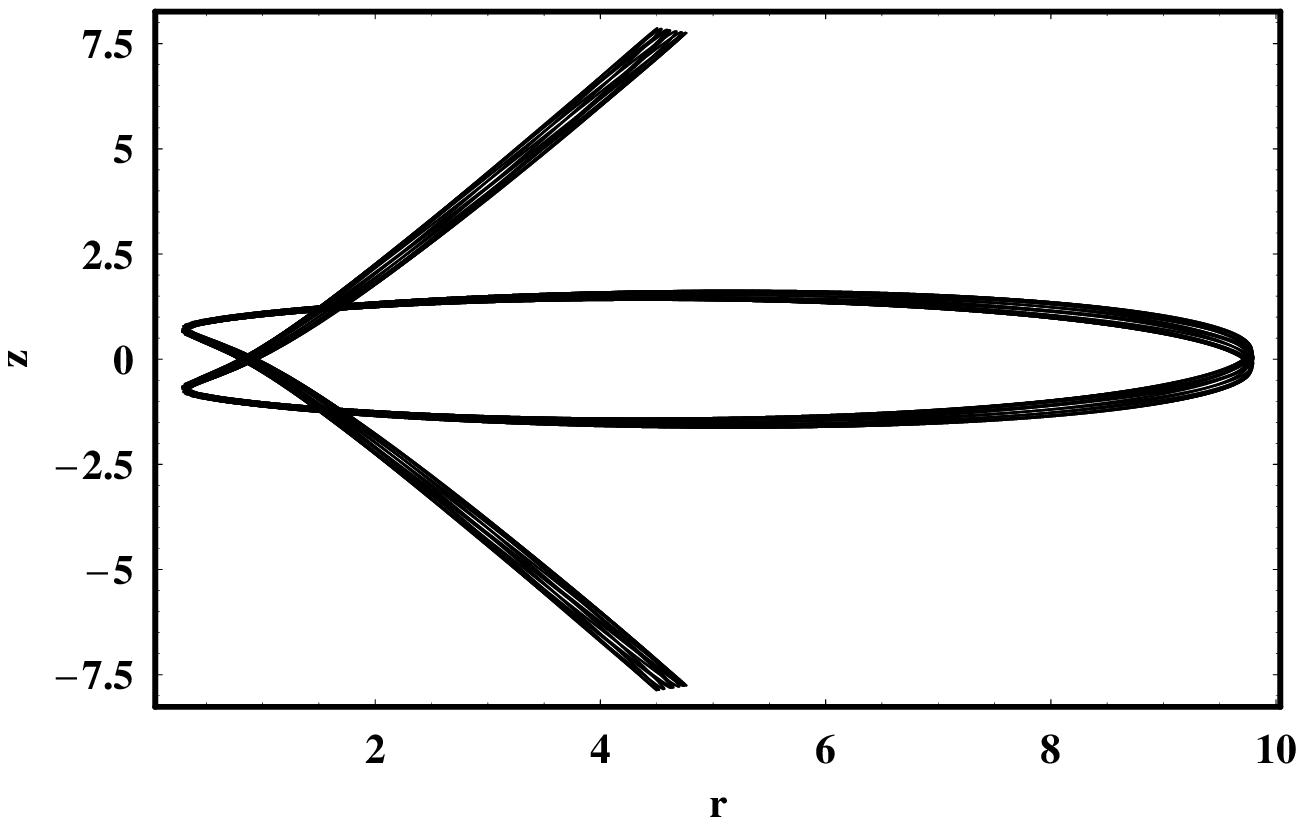}}}
\resizebox{0.8\hsize}{!}{\rotatebox{0}{\includegraphics*{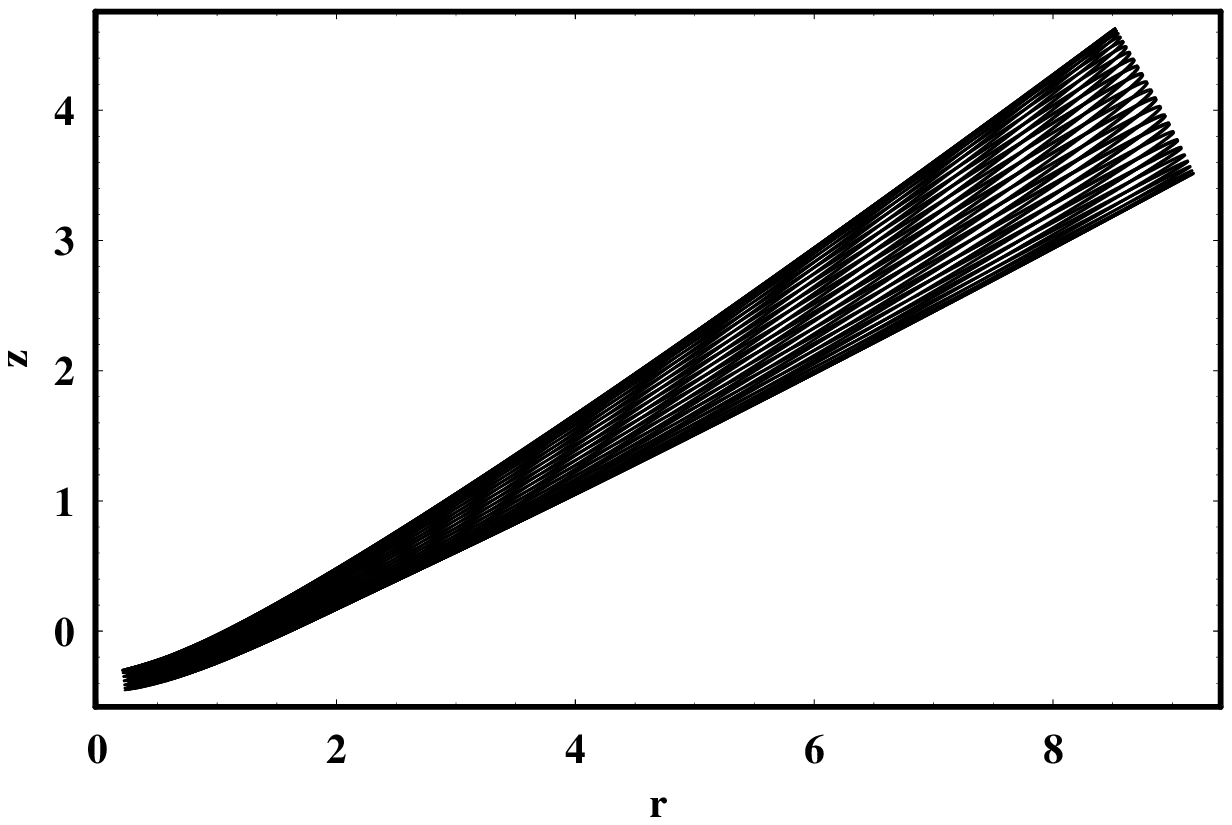}}\hspace{1cm}
                         \rotatebox{0}{\includegraphics*{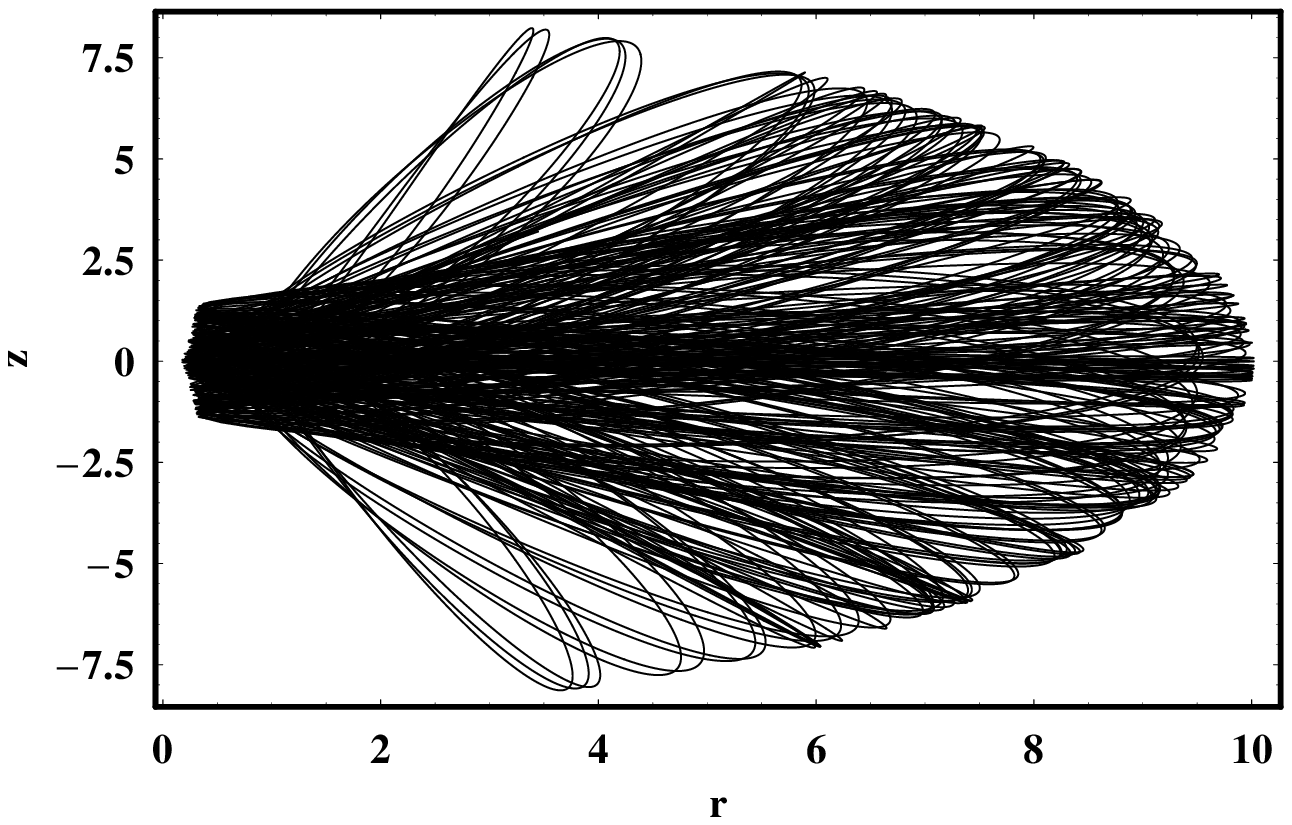}}}
\vskip 0.1cm
\caption{(a-d): Orbits in the elliptical galaxy model, when $E=486$. The values of all parameters are as in Fig. 4.}
\end{figure*}

One can observe large areas of regular motion and a unified chaotic sea. There are three distinct regions of regular orbits. (i) Orbits forming invariant curves surrounding the central invariant point, (ii) orbits forming the triple set of islands and (iii) orbits forming the double set of islands symmetric with respect to the $r$ - axis. The corresponding orbits are shown in Figure 5a-d. Fig 5a shows an orbit forming an island near the central invariant point. The initial conditions are: $r_0=2.8, p_{r0}=0, z_0=0$, while for all orbits the value of $p_{z0}$, is always found using the energy integral (7). This is a typical tube orbit. The orbit shown in Fig. 5b, belongs to group (ii) and its initial conditions are: $r_0=9.8, p_{r0}=0, z_0=0$. This orbit is a characteristic example of the 4/3 resonance. The orbit shown in Fig. 5c belongs to group (iii). The initial conditions of this orbit are: $r_0=1.4, p_{r0}=28.5, z_0=0$. This is a banana orbit and carries stars out of the galactic plane. Finally, the orbit shown in Fig. 5d is a chaotic orbit with initial conditions: $r_0=2.3, p_{r0}=24, z_0=0$. All orbits were calculated for a time period of 100 time units.

\subsection{Disk galaxy model}

Figure 6 shows the structure of the corresponding phase plane when $\upsilon_0=25, a=1.2, b=1.3, h=0.2$. This time we have a disk galaxy. The value of the energy is $E=1425$ and the value of the angular momentum is $L_z=10$.
\begin{figure}[!tH]
\resizebox{\hsize}{!}{\rotatebox{270}{\includegraphics*{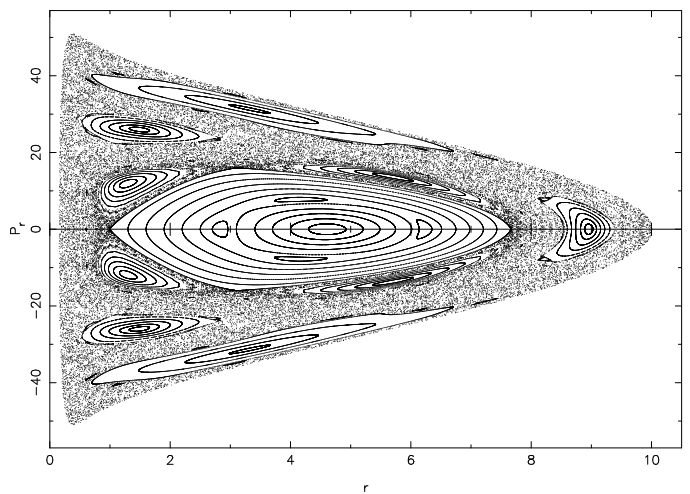}}}
\caption{The $(r-p_r)$ phase plane for the disk galaxy, when $E=1425$. See text for details.}
\end{figure}
\begin{figure*}[!tH]
\centering
\resizebox{0.8\hsize}{!}{\rotatebox{0}{\includegraphics*{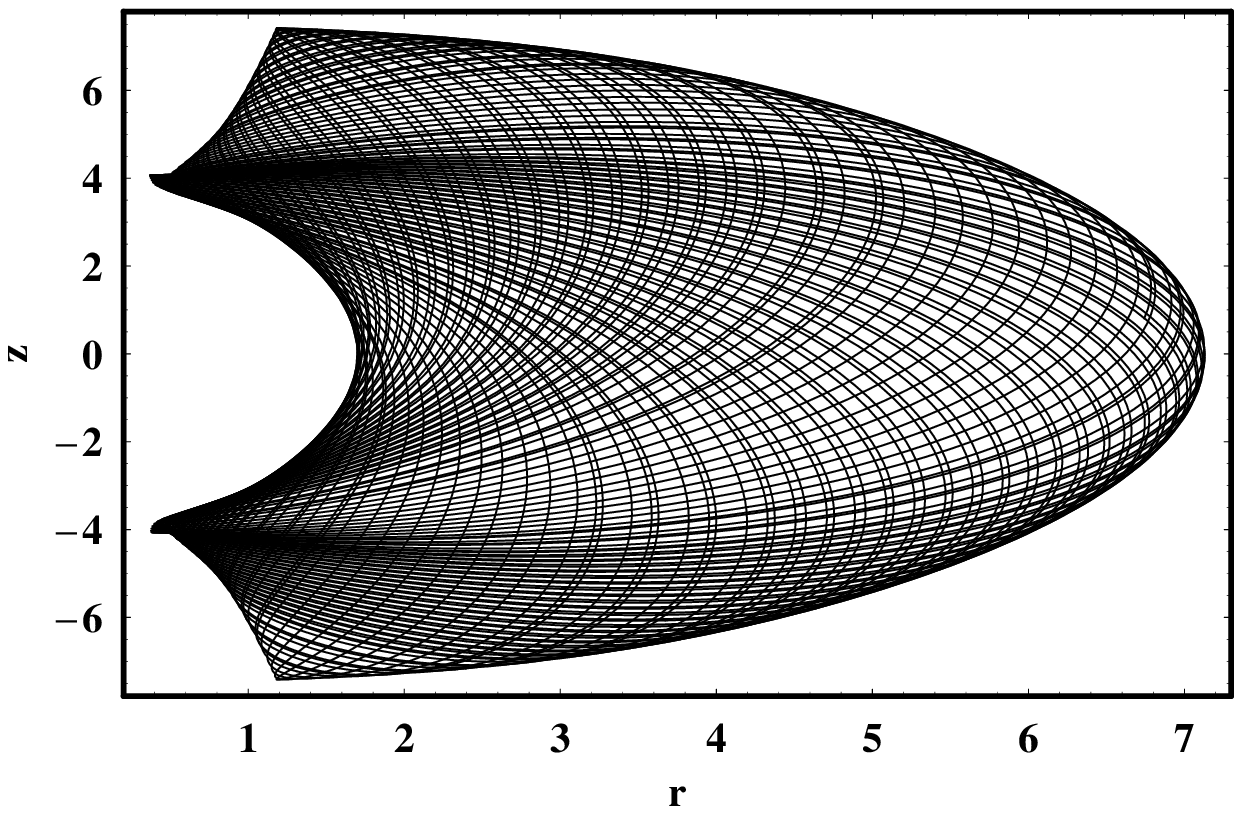}}\hspace{1cm}
                         \rotatebox{0}{\includegraphics*{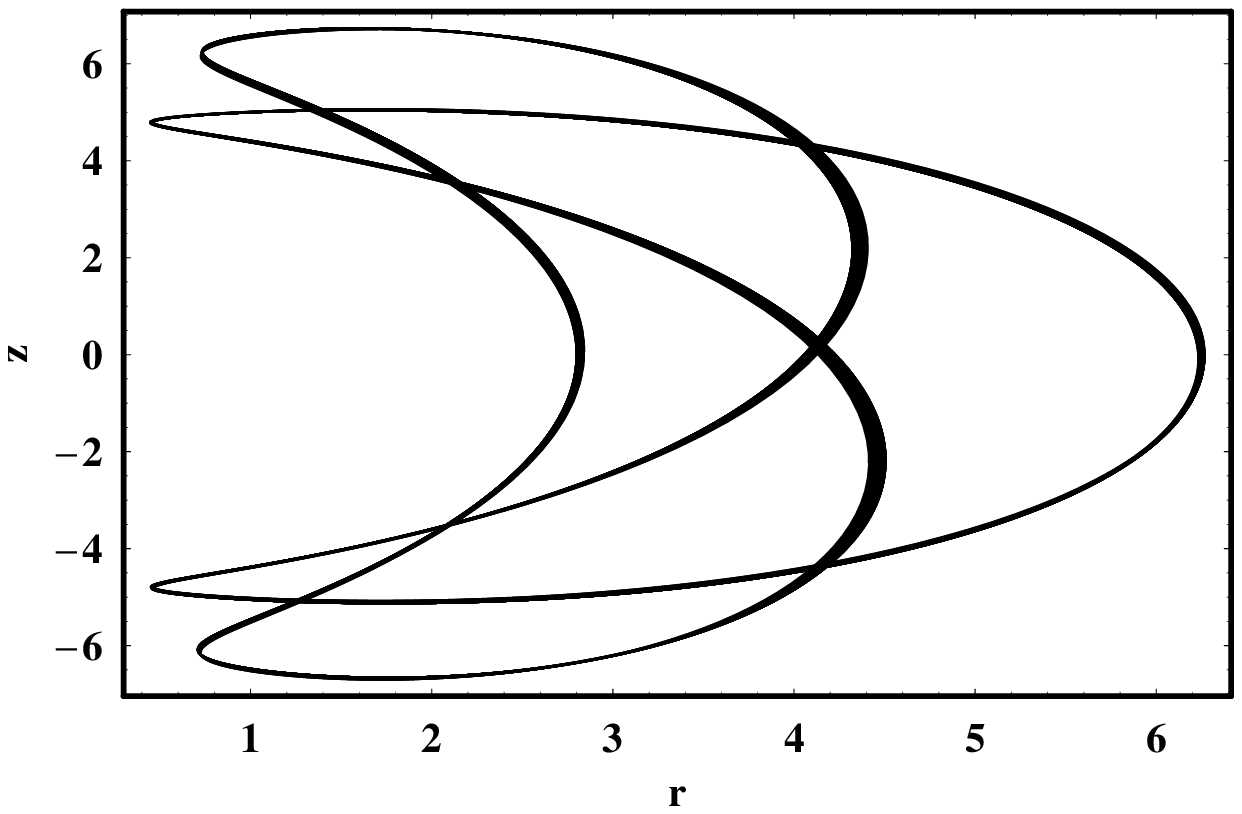}}}
\resizebox{0.8\hsize}{!}{\rotatebox{0}{\includegraphics*{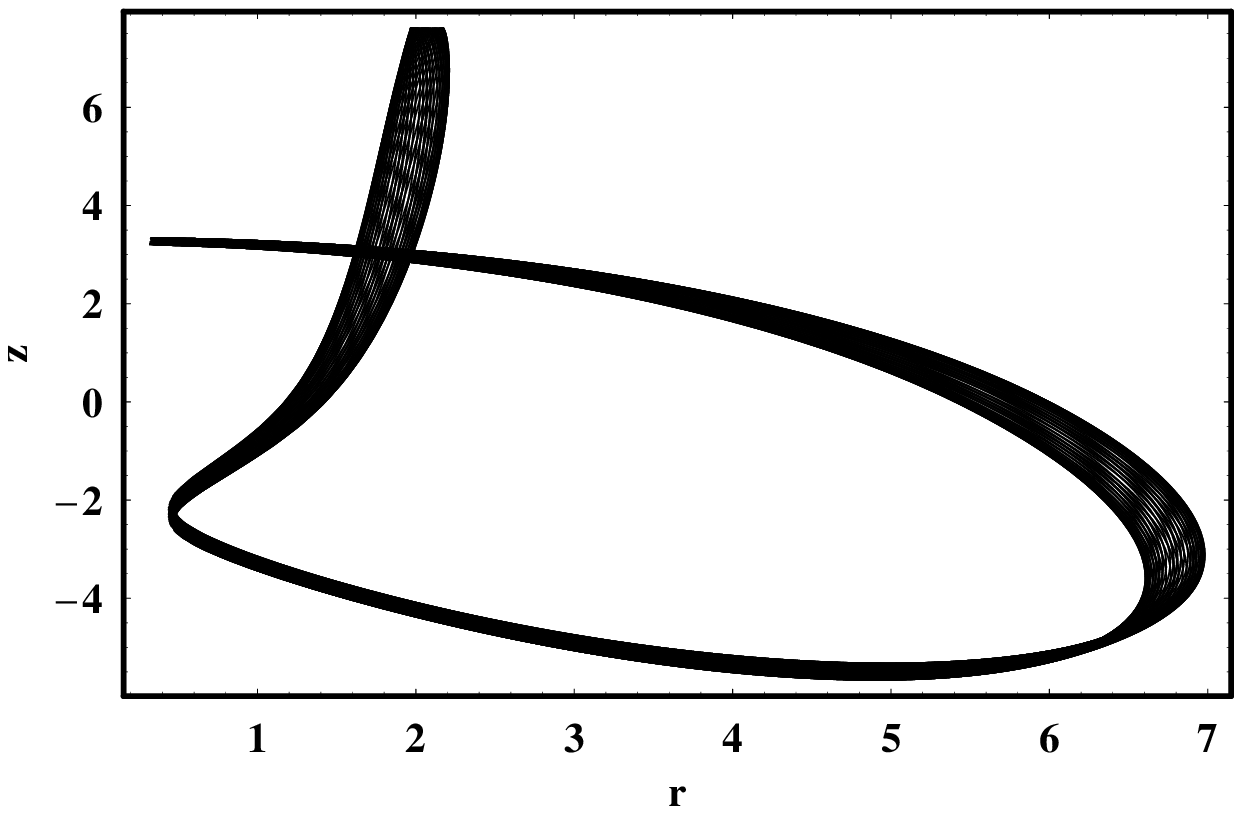}}\hspace{1cm}
                         \rotatebox{0}{\includegraphics*{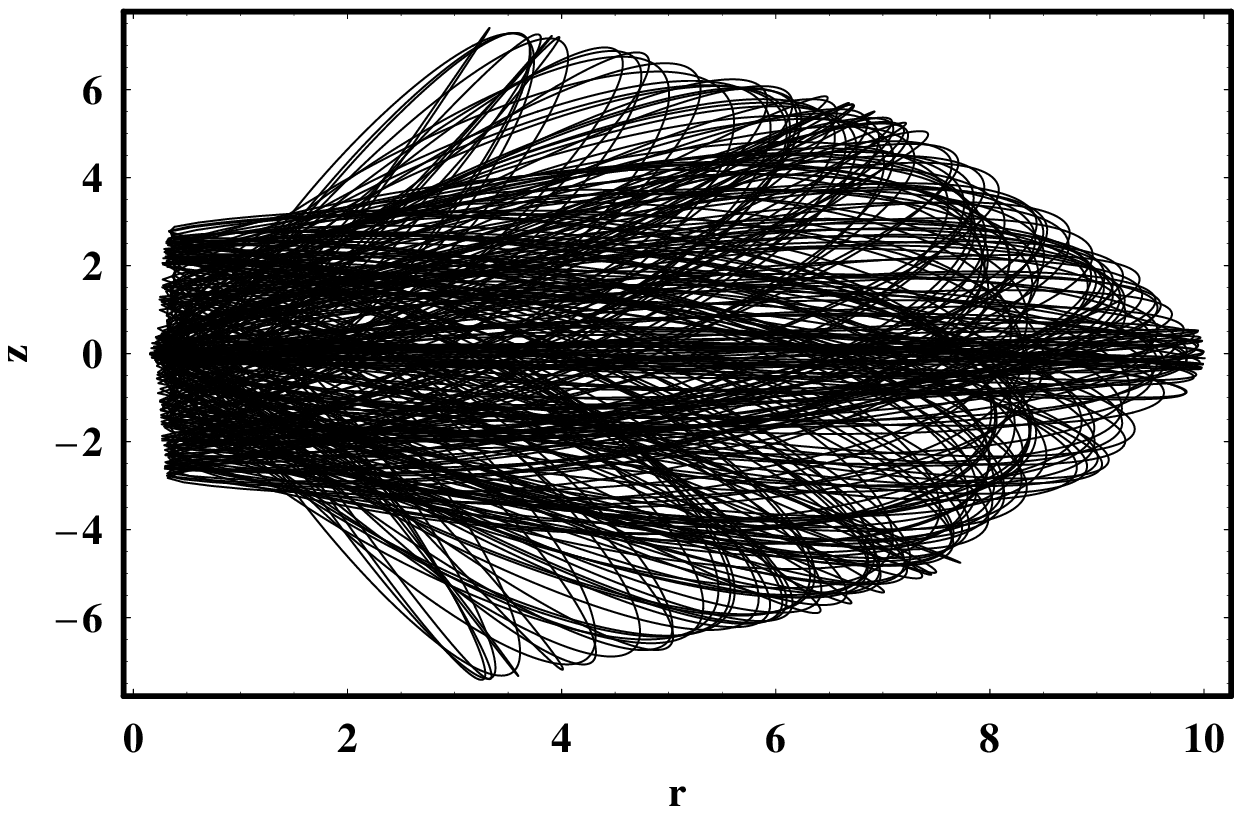}}}
\vskip 0.1cm
\caption{(a-d): Orbits in the disk galaxy model, when $E=1425$. The values of all parameters are as in Fig. 6.}
\end{figure*}

Once more, one can see areas of regular motion and a large chaotic sea. The regular orbits belong to four main families. (i) Orbits forming invariant curves surrounding the central invariant point, (ii) orbits forming the four islands symmetric to the $r$ - axis, (iii) orbits forming the triple set of islands and (iv) orbits forming the outer double set of islands. The corresponding orbits are show in Figure 7a-d. Figure 7a shows an orbit forming an island near the central invariant point. The initial conditions are: $r_0=1.7, p_{r0}=0, z_0=0$. In Fig. 7b we can see a resonant orbit, with initial conditions: $r_0=2.8, p_{r0}=0, z_0=0$. The orbit in Fig. 7c belongs to group (ii). The initial conditions are: $r_0=1.2, p_{r0}=11, z_0=0$. The orbit shown in Fig. 7d is a chaotic orbit with initial conditions: $r_0=0.4, p_{r0}=0, z_0=0$. All orbits were calculated for a time period of 100 time units.
\begin{figure}[!tH]
\centering
\resizebox{0.9\hsize}{!}{\rotatebox{0}{\includegraphics*{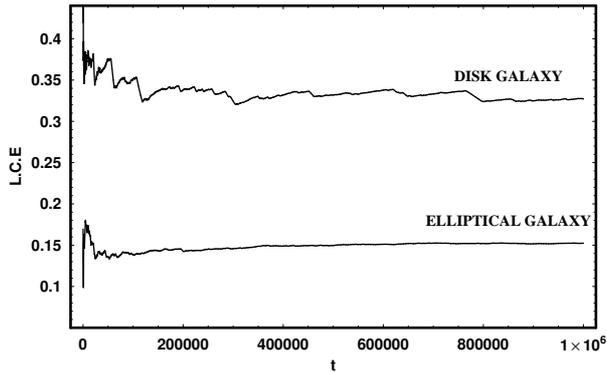}}}
\caption{The L.C.Es for a chaotic orbits in elliptical and disk galaxy models.}
\end{figure}
\begin{figure}[!tH]
\centering
\resizebox{0.9\hsize}{!}{\rotatebox{0}{\includegraphics*{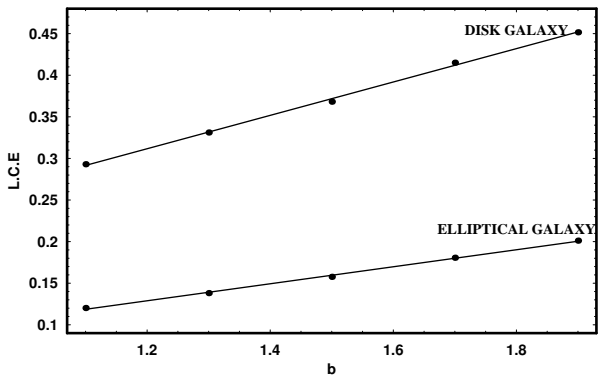}}}
\caption{Maximal L.C.Es of a chaotic orbit in elliptical and disk galaxy vs. b.}
\end{figure}
\begin{figure}[!tH]
\centering
\resizebox{0.9\hsize}{!}{\rotatebox{0}{\includegraphics*{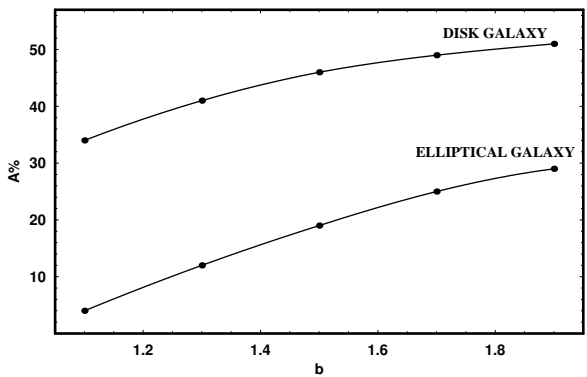}}}
\caption{Plot of the percentage of the area $A\%$ in the phase plane covered by chaotic orbits vs. b for the elliptical and disk galaxy models.}
\end{figure}

\subsection{Comparison of the two models}

It is important to note that in the case of the disk galaxy, the phase plane of the dynamical system displays a wider variety of regular orbits, than in the case of the elliptical galaxy. Moreover, the extent of the chaotic sea in the phase plane of the disk galaxy model (Fig. 6) is significant larger than this of the elliptical galaxy (Fig. 4). In order to have a better estimation of the degree of chaos, we have computed the maximal Lyapunov Characteristic Exponent (L.C.E) (see Lichtenberg and Lieberman, 1992 for details), for a chaotic orbit in each case. The results are shown in Figure 8. The initial conditions of the chaotic orbit for the elliptical galaxy are: $r_0=2.5, p_{r0}=21, z_0=0$, while for the disk galaxy the chaotic orbit has initial conditions: $r_0=9.7, p_{r0}=0, z_0=0$. Both orbits were calculated for a time period of $10^6$ time units. The values of all parameters are as in Fig. 4 (for the elliptical galaxy) and Fig. 6 (for the disk galaxy). It is evident, that the L.C.E of the disk system is twice larger than the L.C.E of the elliptical galaxy. Therefore, one may conclude, that the degree of chaos in disk galaxies is larger, compared to that observed in elliptical galaxies.

Let us now come to see how the flatness parameter $b$ of our model, is connected with other important parameters of the dynamical system. In all cases the value of the flatness parameter $b$, is between $1 < b < 2$. Figure 9 shows a plot of the value of the maximal Lyapunov Characteristic Exponent (L.C.E) of a chaotic orbit in elliptical and disk galaxy vs $b$. One can see, that the values of the L.C.E increases linearly as $b$ increases in both cases. Figure 10 shows the relationship between the percentage of the area $A\%$ covered by chaotic orbits in the phase plane and the flatness parameter $b$. We observe, that in both cases (elliptical and disk galaxy) the percentage $A\%$ increases as the flatness parameter $b$ increases. Since $A\%$ cannot exceed $100\%$, the evolution is slower than a logarithmic.
\begin{figure*}[!tH]
\centering
\resizebox{0.8\hsize}{!}{\rotatebox{0}{\includegraphics*{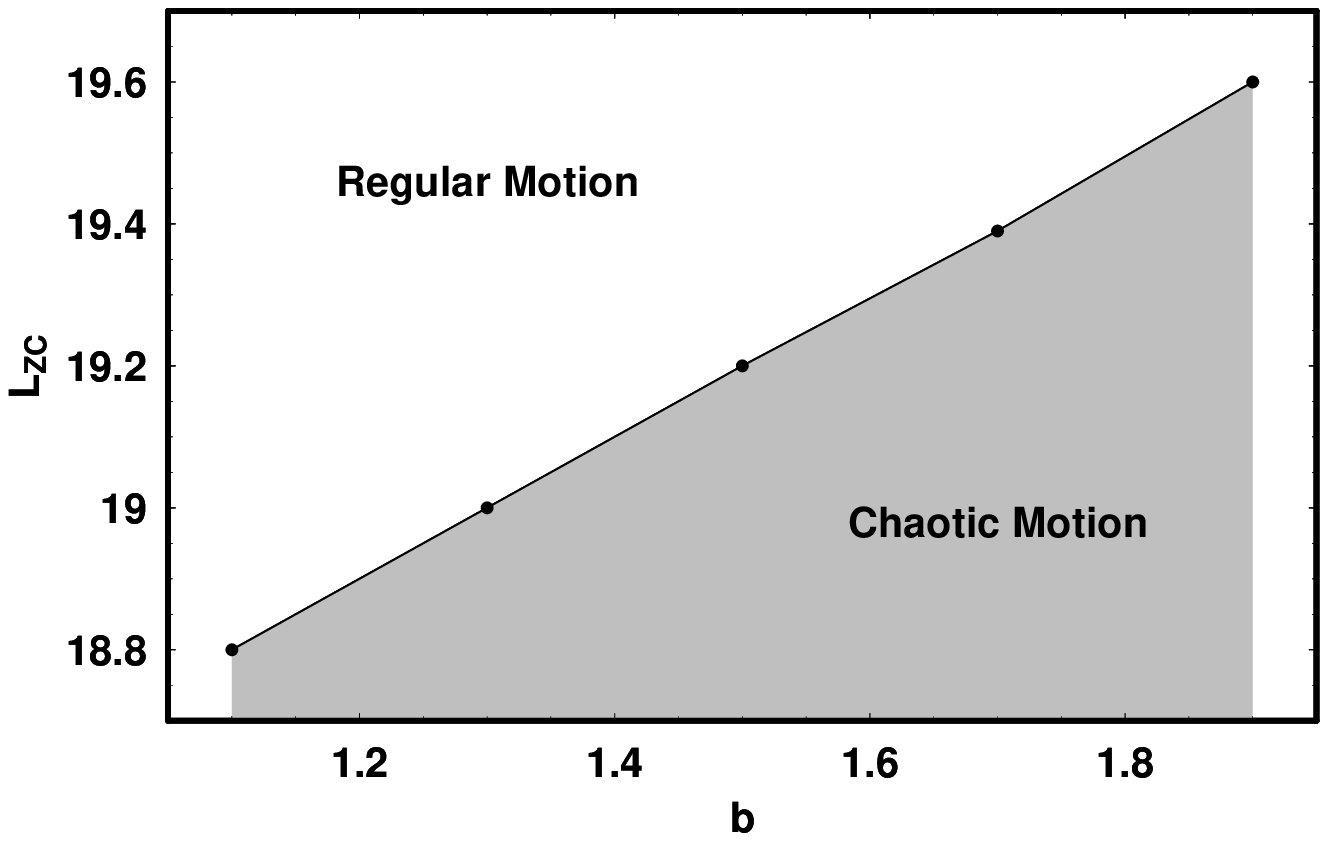}}\hspace{1cm}
                         \rotatebox{0}{\includegraphics*{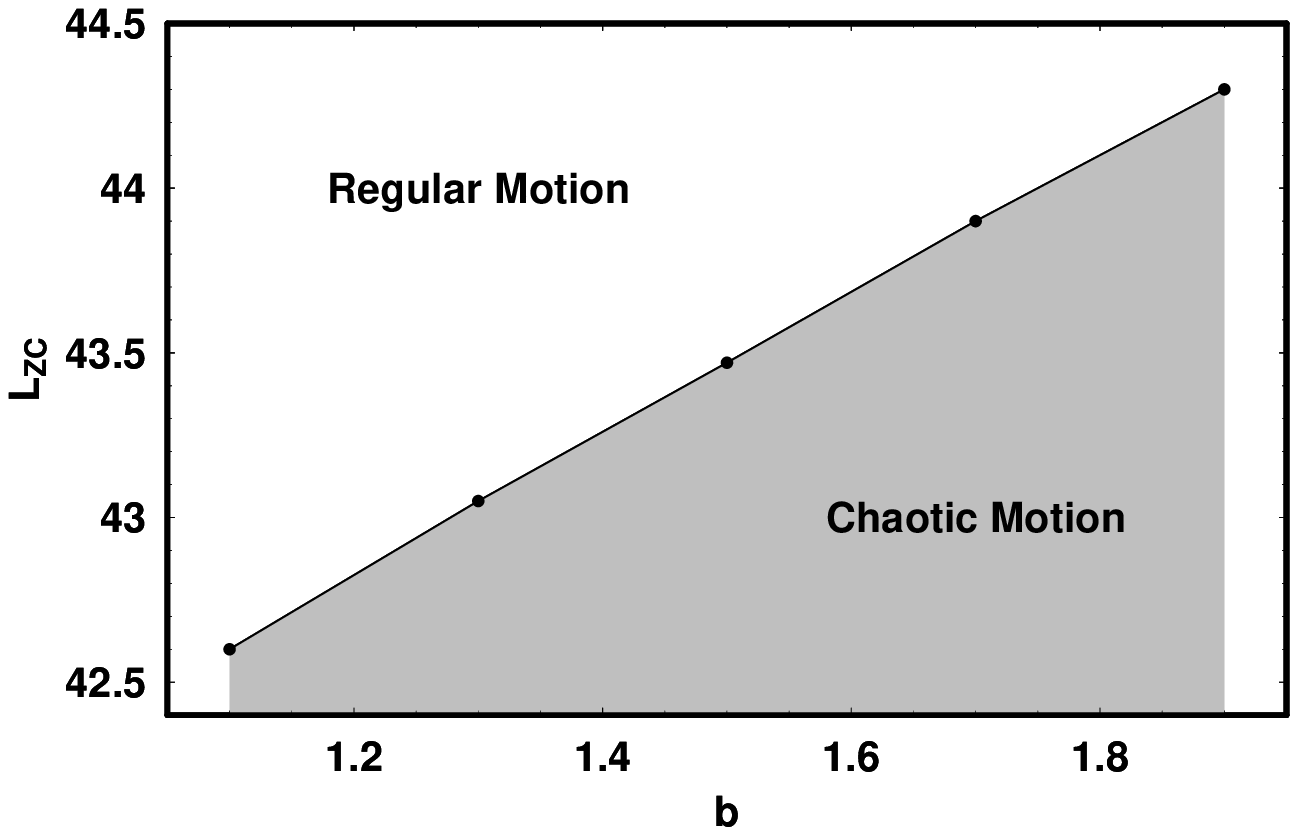}}}
\vskip 0.1cm
\caption{(a-b): Plot of the $L_{zc}$ vs. $b$, (a-left): for the elliptical galaxy and (b-right) for the disk galaxy. See text for details.}
\end{figure*}
\begin{figure}[!tH]
\centering
\resizebox{0.9\hsize}{!}{\rotatebox{0}{\includegraphics*{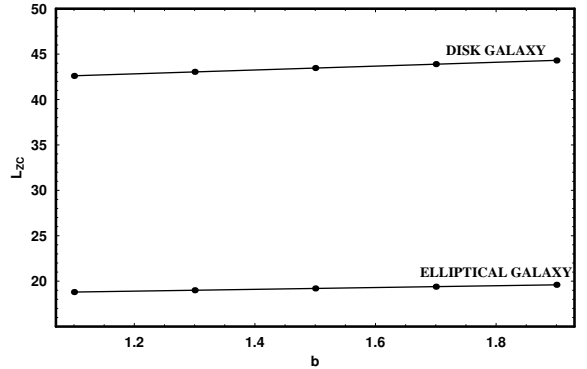}}}
\caption{Plot of the $L_{zc}$ vs. $b$, for both dynamical models.}
\end{figure}

Before closing this Section, we would like to present a relationship between the critical value of the angular momentum $L_{zc}$ (that is the maximum value of the angular momentum, for which, stars moving near the galactic plane are scattered, displaying chaotic motion, for a given value of the flatness parameter $b$), when all other parameters are kept constant. The results, which were found numerically, are given in Figures 11a-b. For the elliptical galaxy of Fig. 11a the values of all the parameters are as in Fig. 4, while for the disk galaxy of Fig. 11b the values of all the parameters are as in Fig. 6. Orbits starting in the upper part of the $\left[b-L_{zc}\right]$ planes are regular, while orbits starting in the lower part of this plane including the line are chaotic. We see in Figure 12, that in both cases the relationship between the flatness parameter $b$ and $L_{zc}$ in linear. An explanation of all the above relationships, will be given in Section 5.
\begin{figure*}[!tH]
\centering
\resizebox{0.8\hsize}{!}{\rotatebox{0}{\includegraphics*{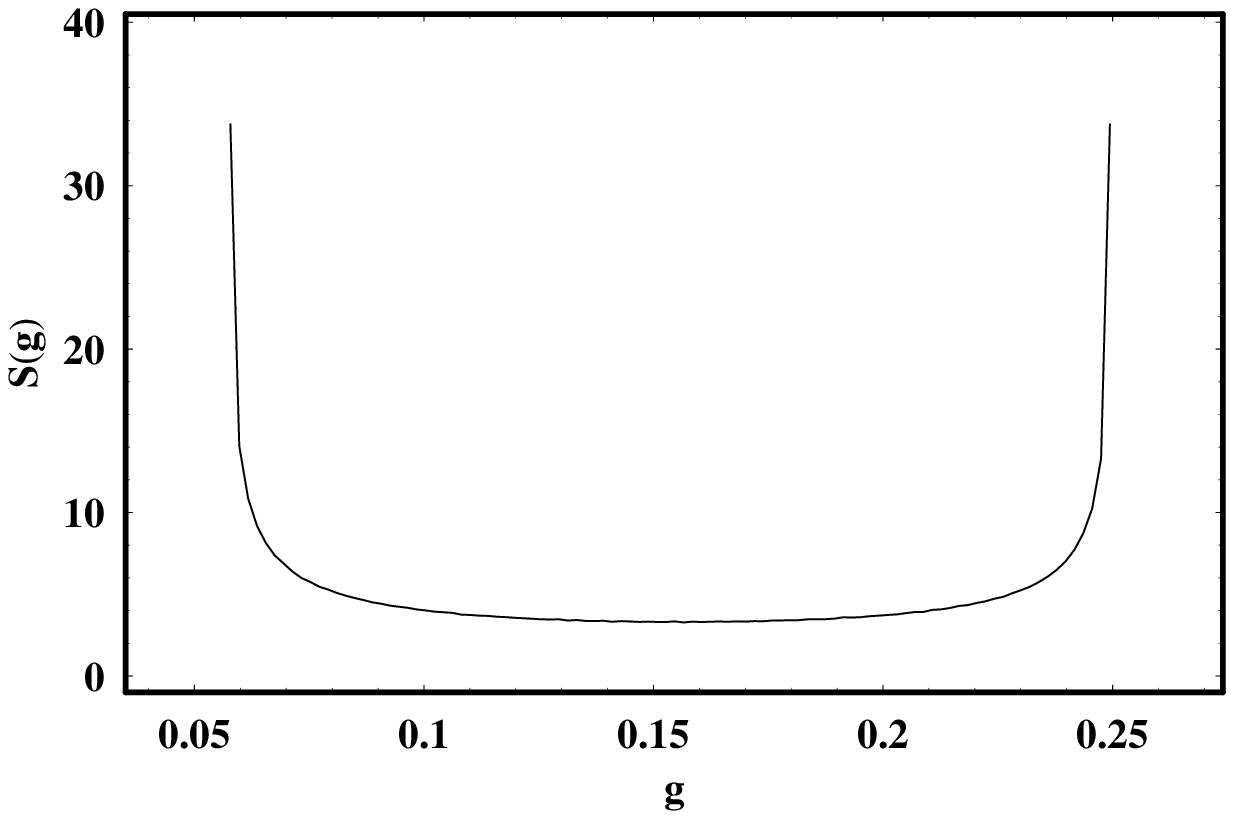}}\hspace{1cm}
                         \rotatebox{0}{\includegraphics*{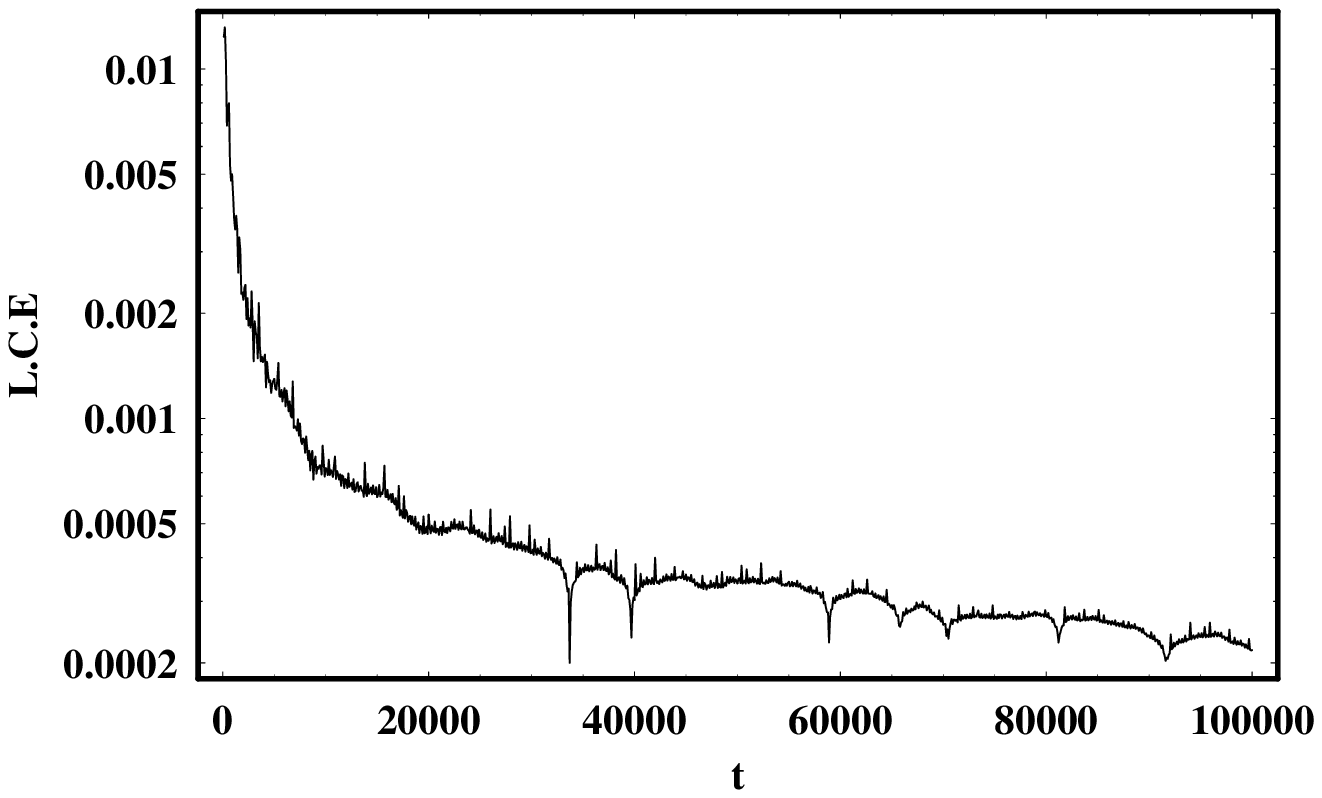}}}
\vskip 0.1cm
\caption{(a-b): (a-left): The $S(g)$ spectrum for a regular orbit and (b-right): The corresponding L.C.E.}
\end{figure*}
\begin{figure*}[!tH]
\centering
\resizebox{0.8\hsize}{!}{\rotatebox{0}{\includegraphics*{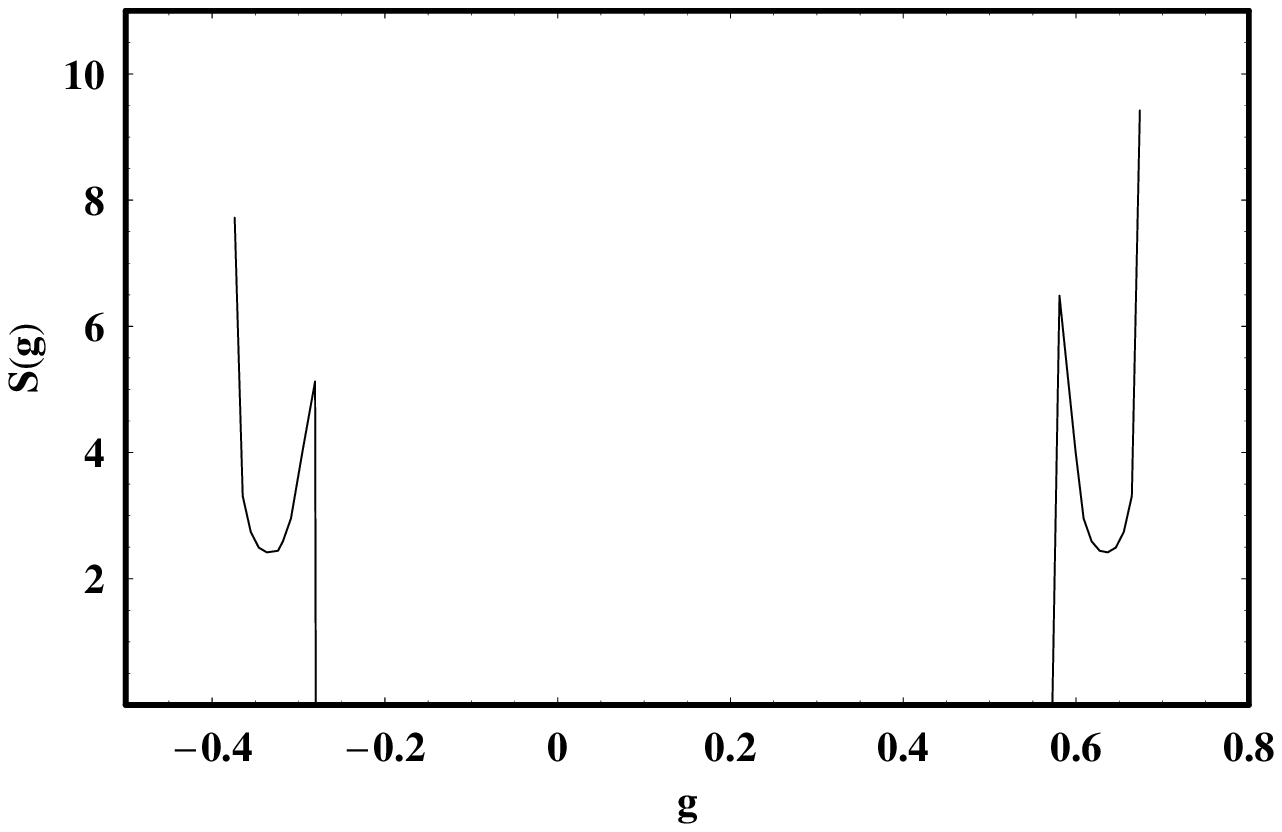}}\hspace{1cm}
                         \rotatebox{0}{\includegraphics*{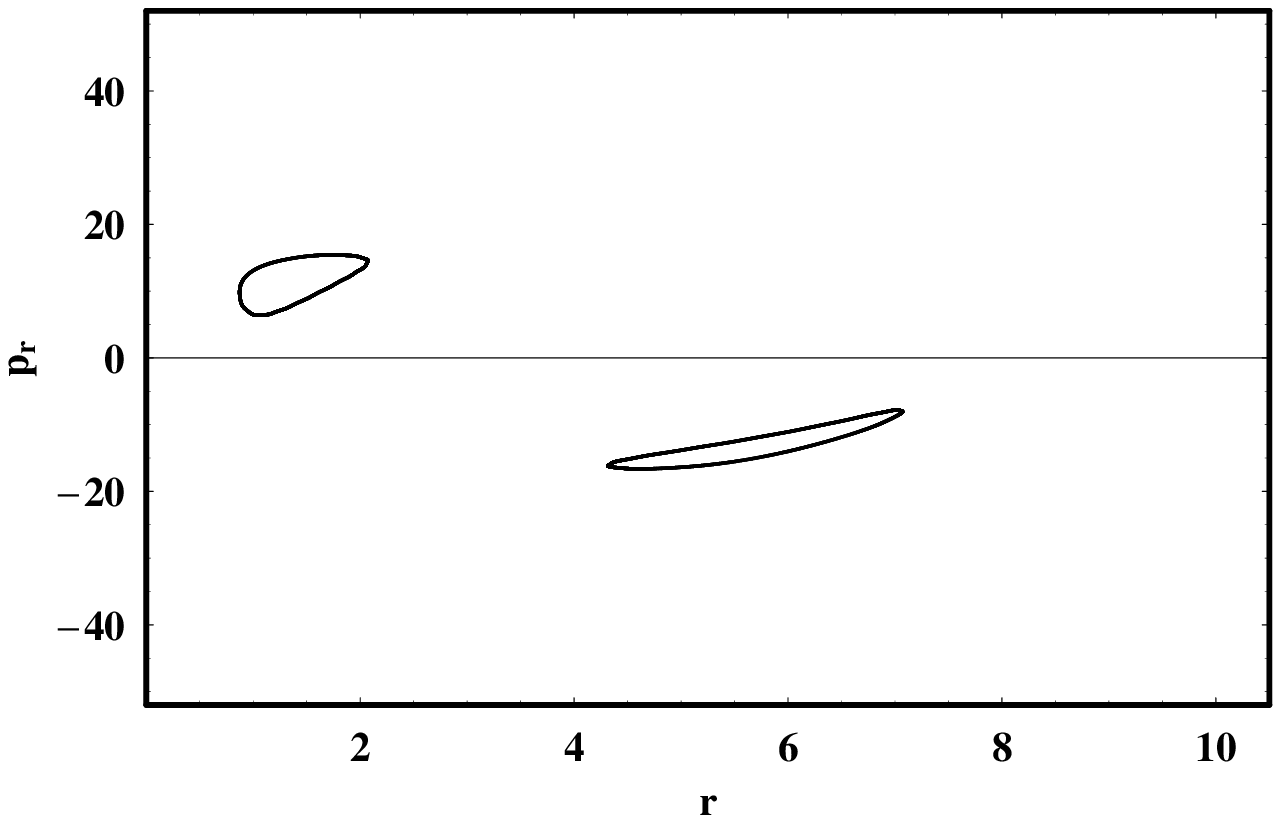}}}
\vskip 0.1cm
\caption{(a-b): The $S(g)$ spectrum for an orbit producing two islands and (b-right): The two islands in the $(r-p_r)$ phase plane of Fig. 6.}
\end{figure*}
\begin{figure*}[!tH]
\centering
\resizebox{0.8\hsize}{!}{\rotatebox{0}{\includegraphics*{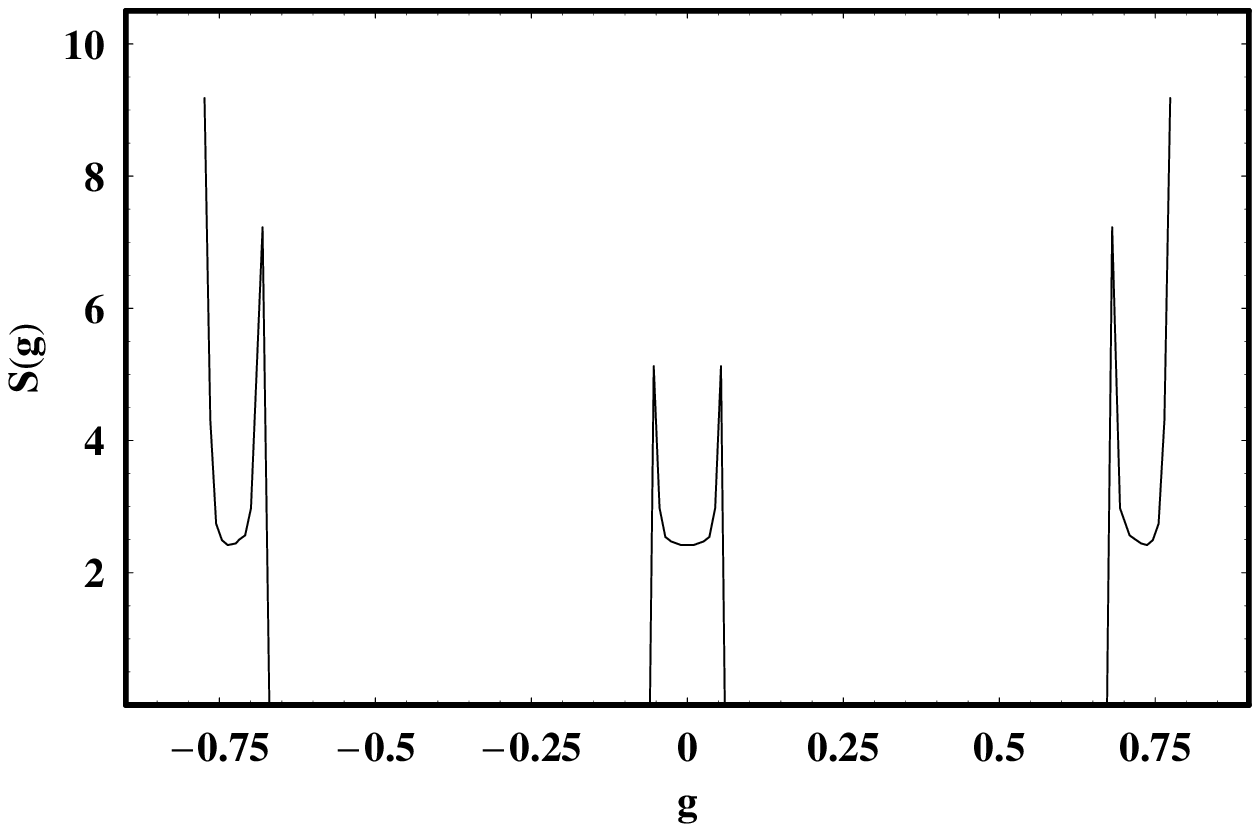}}\hspace{1cm}
                         \rotatebox{0}{\includegraphics*{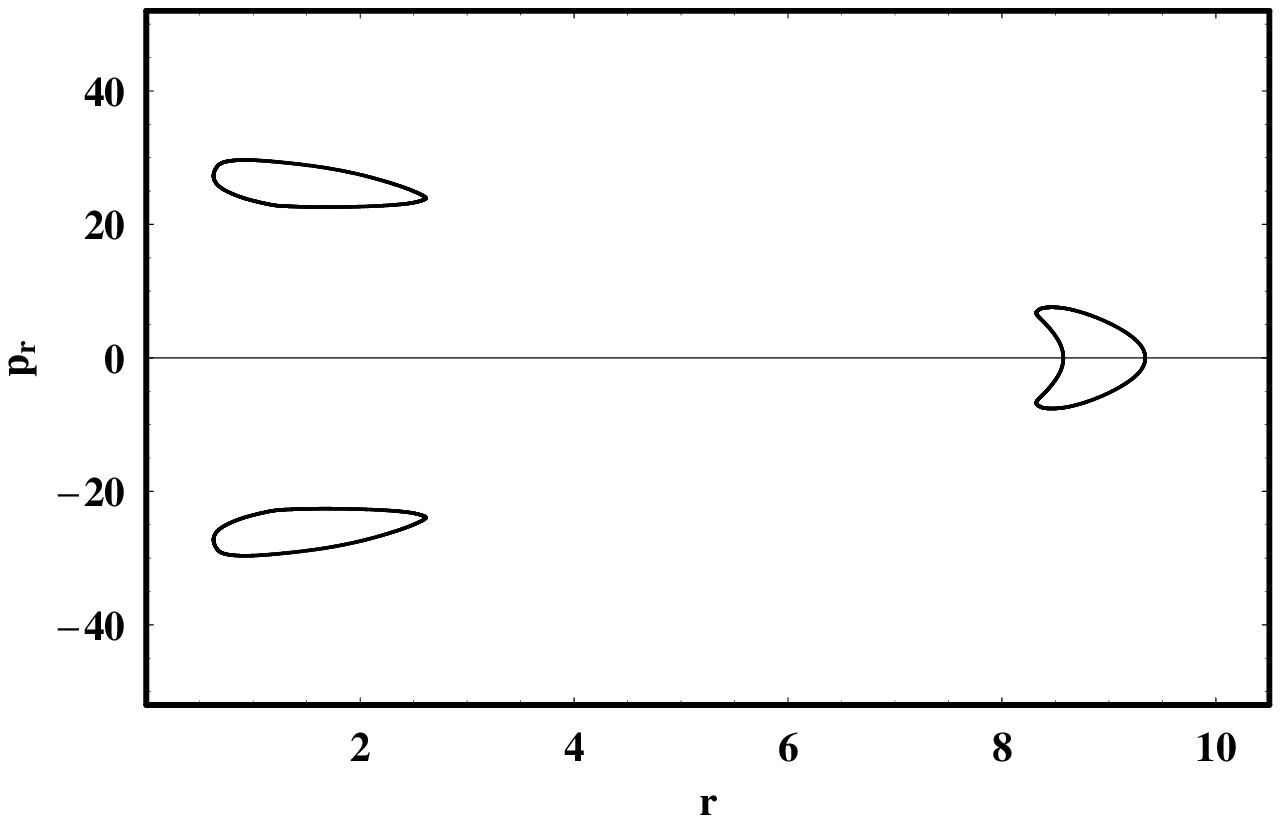}}}
\vskip 0.1cm
\caption{(a-b): The $S(g)$ spectrum for an orbit producing three islands and (b-right): The three islands in the $(r-p_r)$ phase plane of Fig. 6.}
\end{figure*}
\begin{figure*}[!tH]
\centering
\resizebox{0.8\hsize}{!}{\rotatebox{0}{\includegraphics*{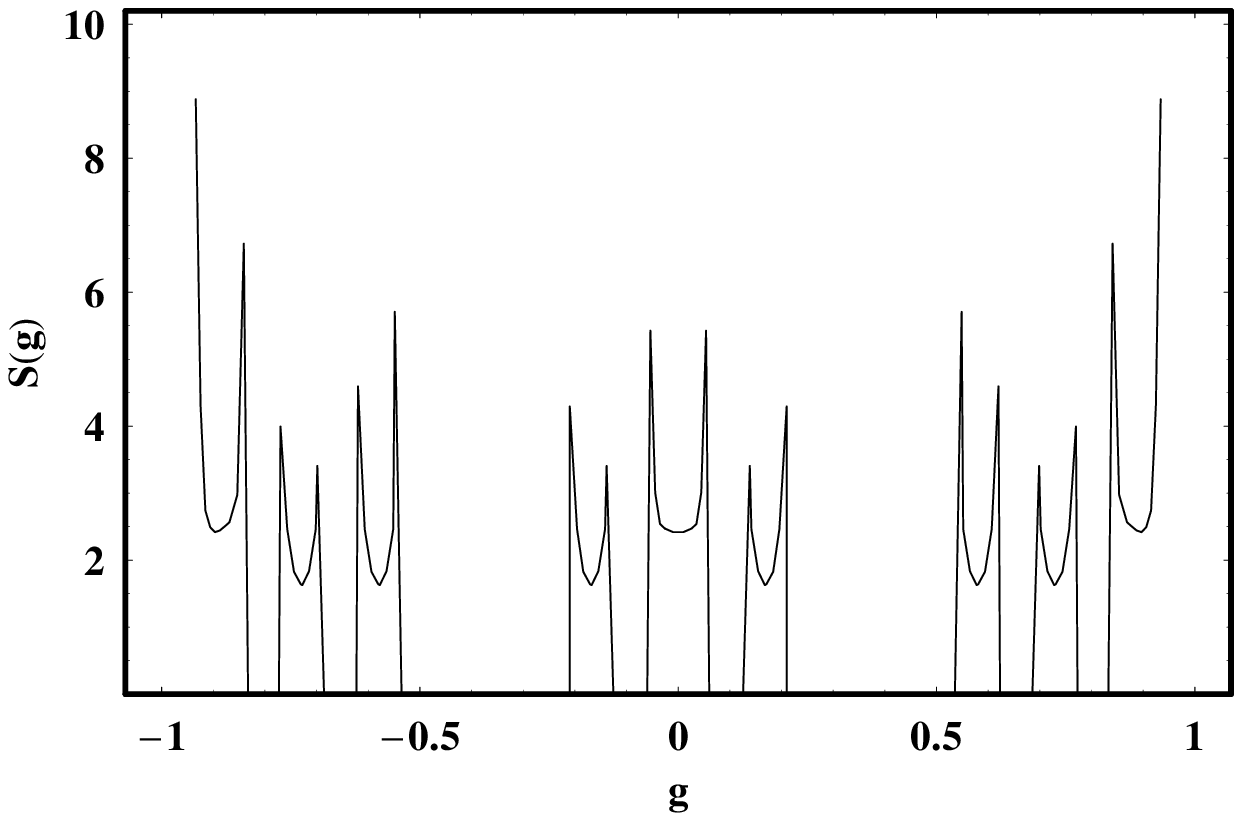}}\hspace{1cm}
                         \rotatebox{0}{\includegraphics*{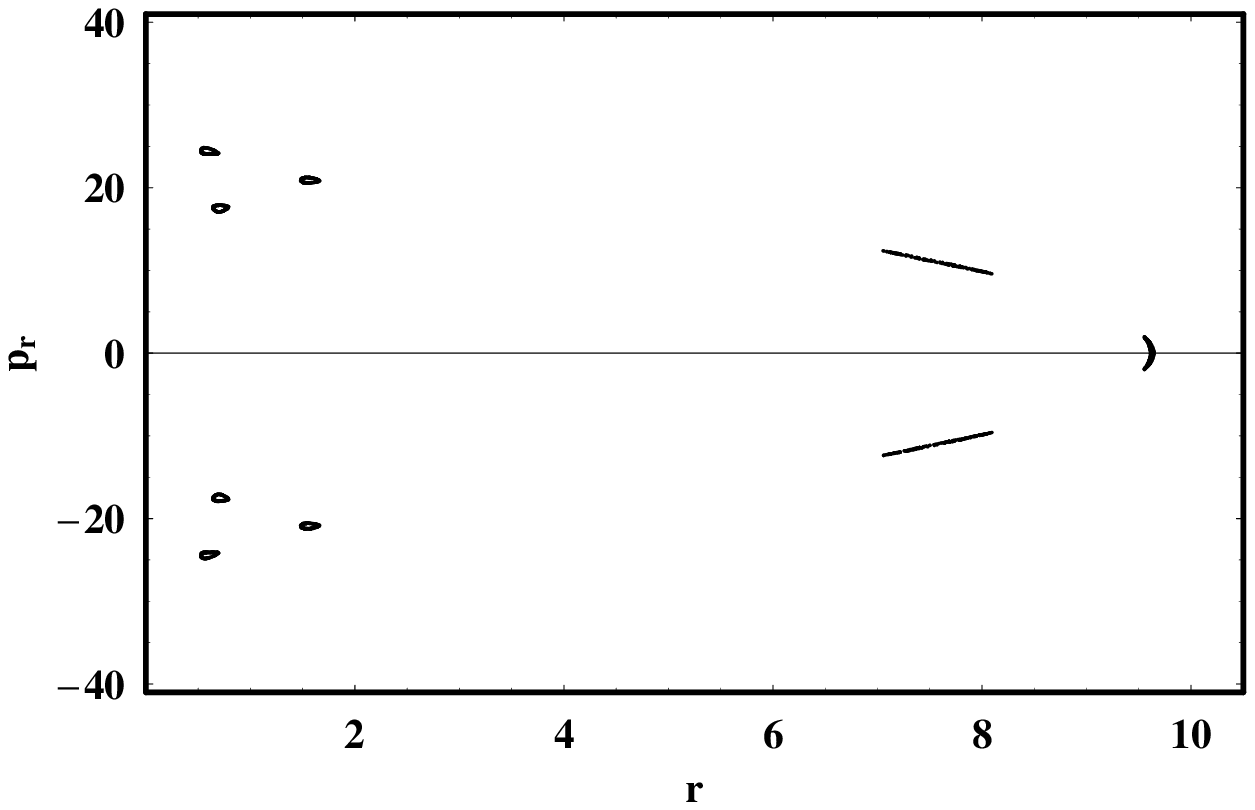}}}
\vskip 0.1cm
\caption{(a-b): Similar to Fig. 14a-b, but for an orbit producing nine small islands in the $(r-p_r)$ phase plane of Fig. 4.}
\end{figure*}

\section{The S(g) dynamical spectrum}

During the last twenty years, dynamical spectra have been frequently used in Galactic Dynamics, as fast indicators of regular or chaotic motion (see Contopoulos et al. 1995, Contopoulos and Voglis 1996, Patsis et al. 1997, Caranicolas \& Vozikis 1999, Karanis \& Caranicolas 2001, Caranicolas \& Papadopoulos 2007, Caranicolas \& Zotos 2010). In this Section, we shall introduce and use, a new type of dynamical spectrum, which is the $S(g)$ spectrum.

We define the dynamical parameter $g$ as
\begin{equation}
g_i=\frac{r_i+p_{ri}-r_ip_{ri}}{p_{zi}},
\end{equation}
where $r_i,p_{ri},p_{zi}$ are the successive values on the Poincar\'{e} $(r,p_r)$, $z=0, p_z>0$, surface of section. We shall define the new dynamical spectrum of the parameter $g$, through its distribution function
\begin{equation}
S(g)=\frac{\Delta N(g)}{N\Delta g},
\end{equation}
where $\Delta N(g)$ are the number of the parameters $g$ in the interval $(g, g+\Delta g)$, after $N$ iterations. The reader can find more details and theoretical explanations, regarding the shapes and behaviors of the dynamical spectra, in the pioneer works of Froeschl\'{e} et al. 1993, Voglis and Contopoulos 1994 and Contopoulos et al. 1997. By definition, the $g$ parameter is based on a complicate combination of coordinates and momenta. The numerator of the parameter $g$ ia a function of $r$ and $p_r$ while, the denominator is a much more complex function of the same variables, since $p_z$ is related to $r$ and $p_r$ through the energy integral (7). Due to the fact that the values of $p_{zi}$ are the successive values on the Poincar\'{e} surface of section, they can never obtain a zero value (always $p_z>0$), which can infinite the $g$ parameter. There are many reasons for introducing the new $S(g)$ spectrum. Three important reasons are that: (i) it can identify islandic motion of resonant orbits, (ii) it can help us to understand the evolution of sticky motion and (iii) it is much faster than the L.C.E, because it needs about $10^4$ time units, while L.C.E needs about $10^5$ time units, in order to obtain reliable results.
\begin{figure*}[!tH]
\centering
\resizebox{0.8\hsize}{!}{\rotatebox{0}{\includegraphics*{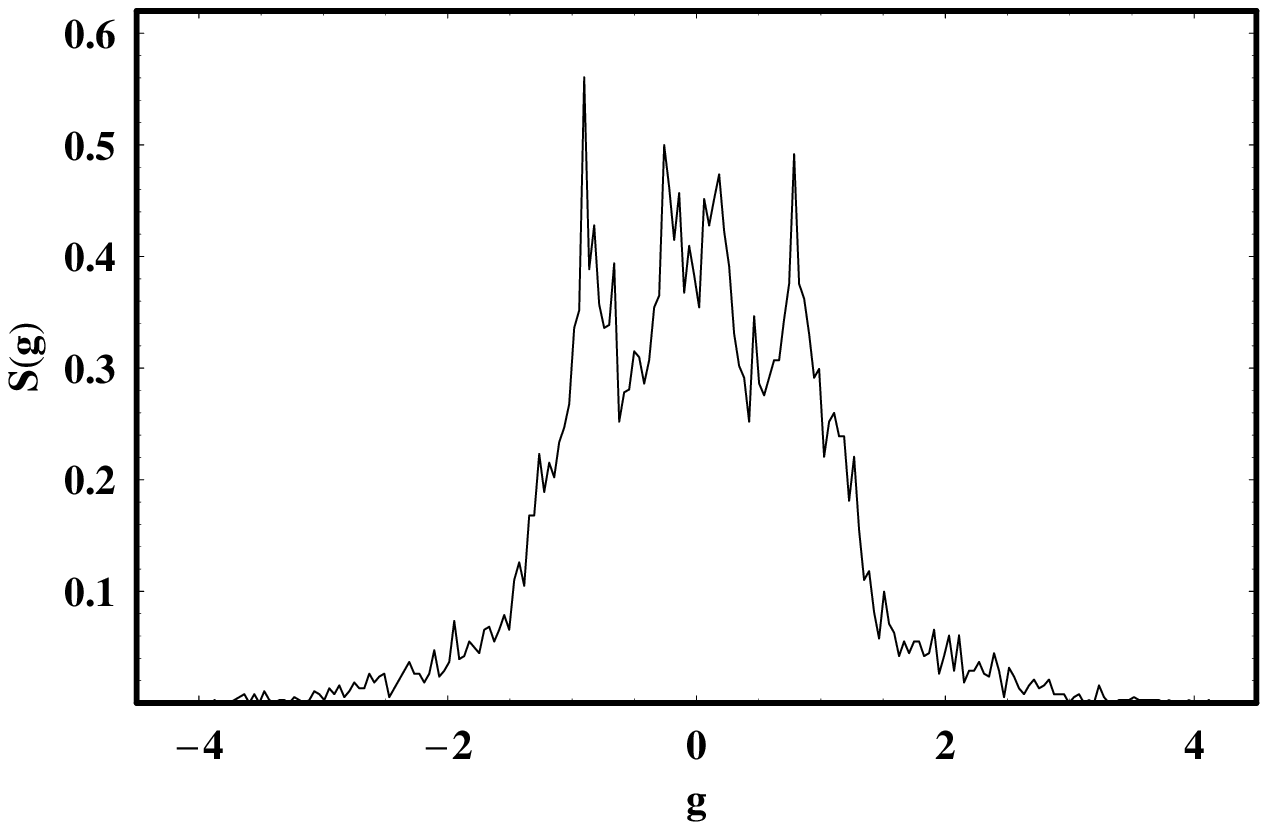}}\hspace{1cm}
                         \rotatebox{0}{\includegraphics*{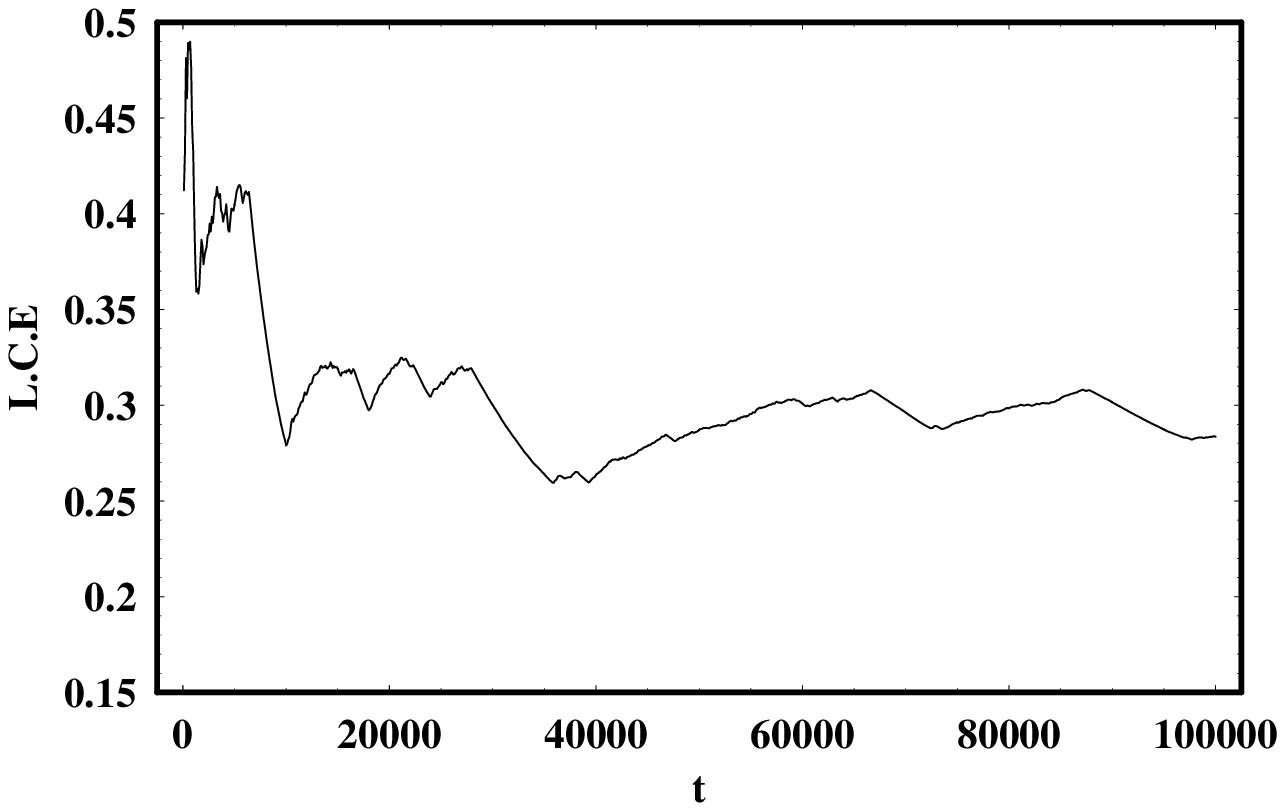}}}
\vskip 0.1cm
\caption{(a-b): (a-left): The $S(g)$ spectrum for a chaotic orbit and (b-right): The corresponding L.C.E.}
\end{figure*}

Figure 13a shows the $S(g)$ spectrum for an orbit, producing one of the invariant curves surrounding the central fixed invariant point shown in the phase plane of Fig. 6. Initial conditions are: $r_0=4, p_{r0}=0, z_0=0$, while the value of $p_{z0}$ is always found from the energy integral (7). As expected, one can observe a well defined $U$ type spectrum, indicating regular motion. The spectrum shown in Fig. 13a is entirely in the $g>0$ domain. This means that the corresponding orbit is a typical orbit of the 1:1 resonance, forming an invariant curve surrounding the central invariant point of Fig. 6. The corresponding L.C.E for a time period of $10^5$ time units, is given in Fig. 13b. Figure 14a shows the $S(g)$ spectrum producing two islands shown in Fig. 6, with initial conditions: $r_0=1, p_{r0}=6.5, z_0=0$, while in Fig. 14b we see the two islands in the $(r-p_r)$ phase plane. Figure 15a shows the $S(g)$ spectrum, for an orbit producing three islands, shown in Fig. 6. Initial conditions are: $r_0=9.34, p_{r0}=0, z_0=0$. Here we observe three $U$ type spectra, which is as much the number of islands of Fig. 15b. Moreover, we see that the left and right spectra are exactly symmetrical about the $g=0$ axis, while the central spectrum lies on both sides of this axis. This indicates, that two of the islands are symmetrical about the $r$ - axis, while the third intersects the $r$ - axis. In other words, we have the case of a quasi-periodic orbit, with a starting point on the $r$ - axis. Figures 16a-b are similar to Figs. 15a-b, but for an orbit producing nine small islands shown in Fig. 4. Initial conditions are: $r_0=9.613, p_{r0}=0, z_0=0$.
\begin{figure*}[!tH]
\centering
\resizebox{0.8\hsize}{!}{\rotatebox{0}{\includegraphics*{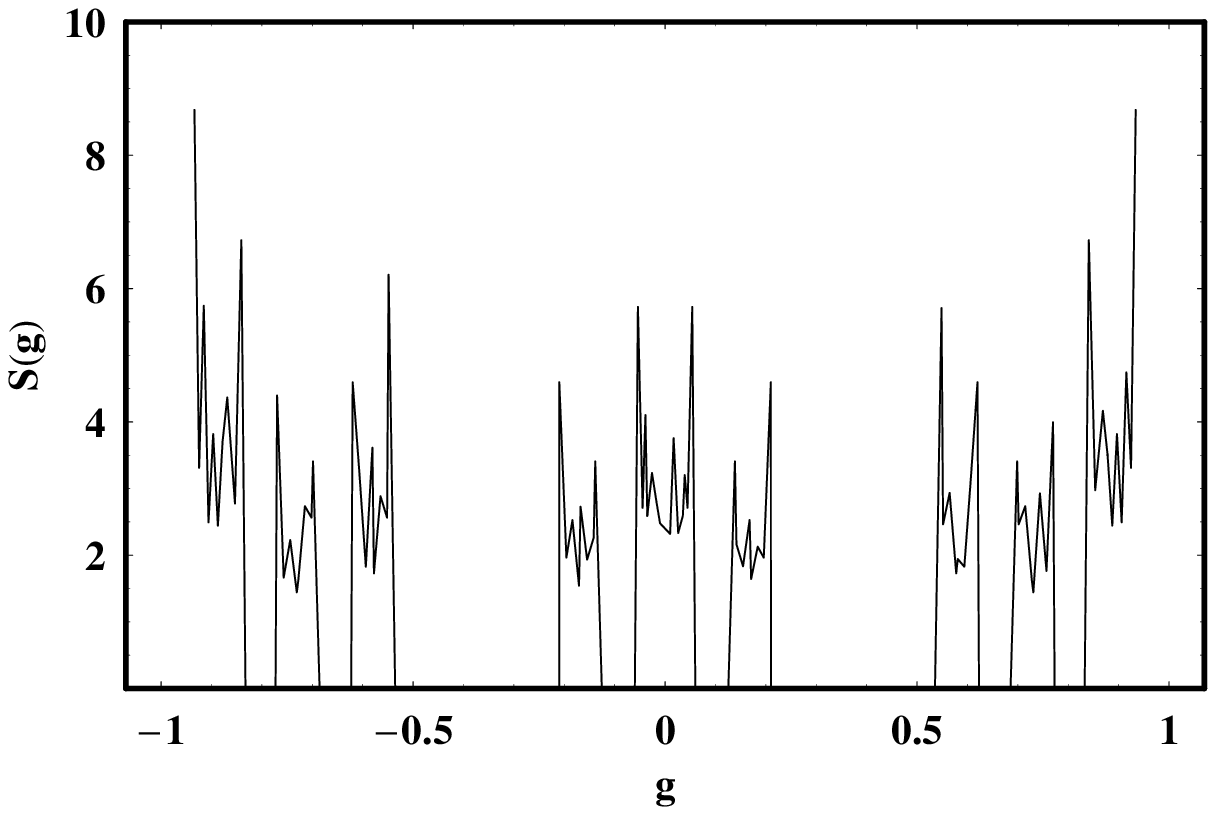}}\hspace{1cm}
                         \rotatebox{0}{\includegraphics*{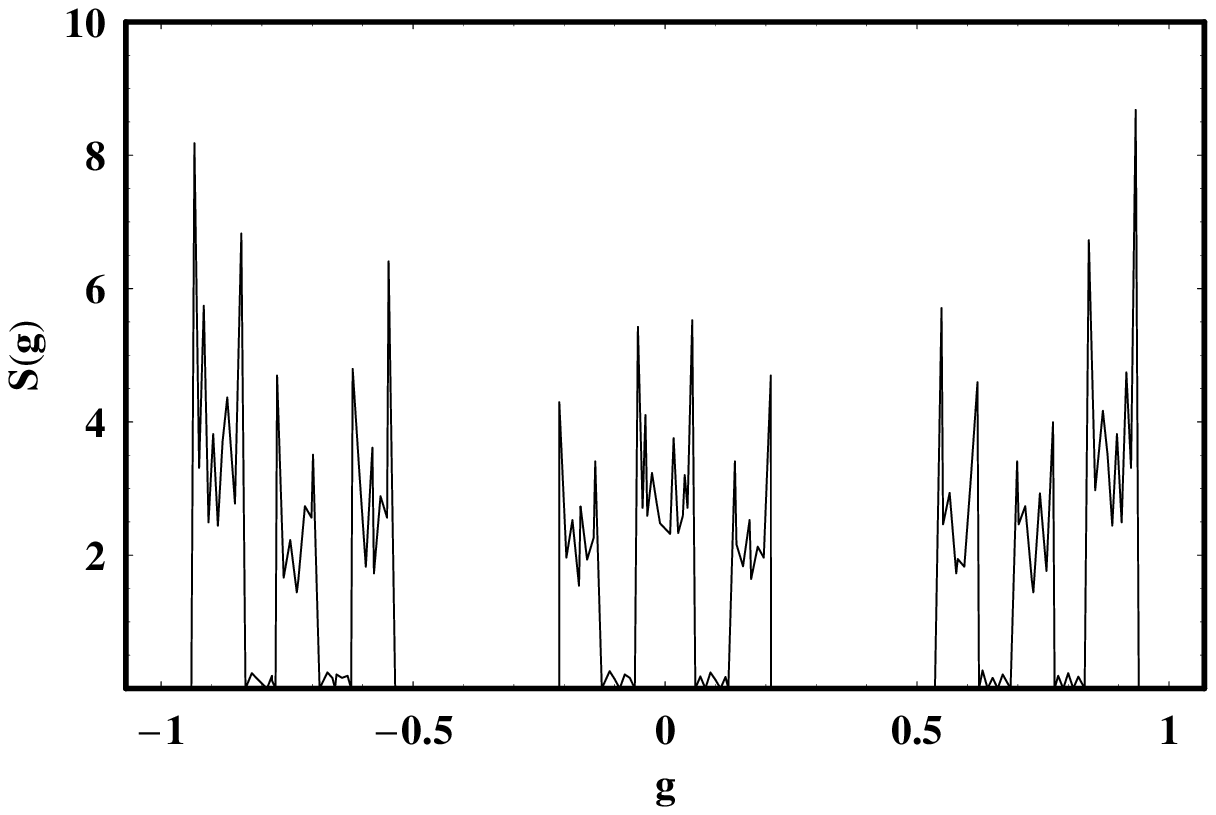}}}
\resizebox{0.8\hsize}{!}{\rotatebox{0}{\includegraphics*{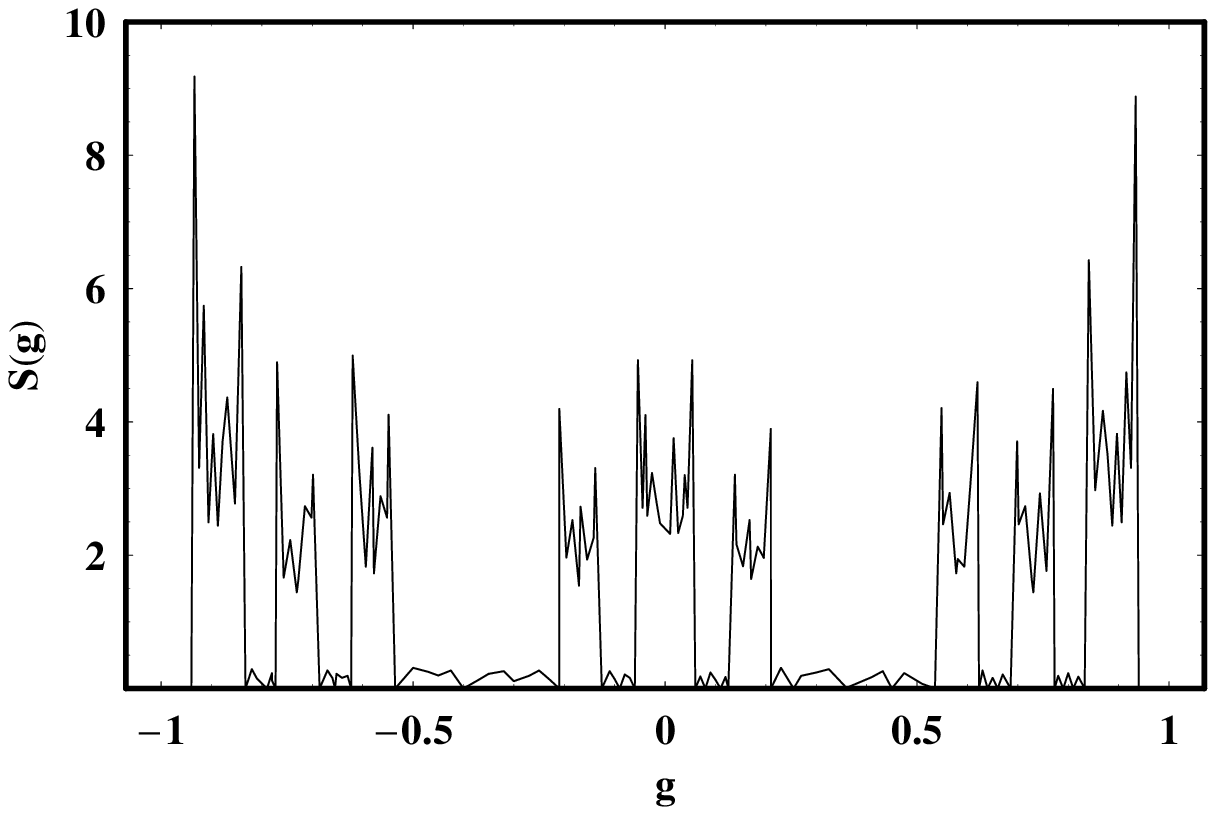}}\hspace{1cm}
                         \rotatebox{0}{\includegraphics*{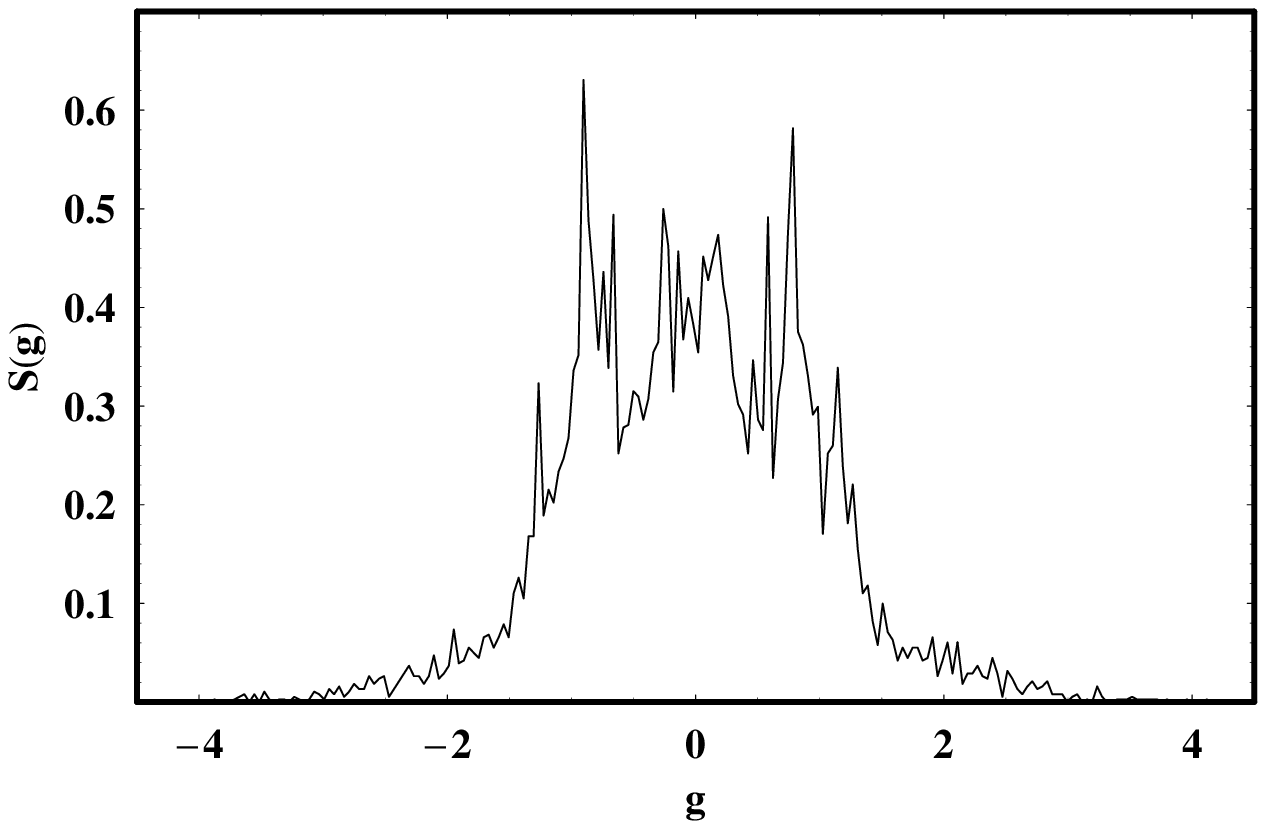}}}
\vskip 0.1cm
\caption{(a-d): Evolution of the $S(g)$ spectrum of a sticky orbit. Details are given in text.}
\end{figure*}

Thus, we came to the conclusion that the $S(g)$ spectrum is very useful, in order to identify islandic motion of resonant orbits. In this point, we should emphasize that, with some of previously used spectra (see Karanis \& Caranicolas 2002 and references therein) we were able to detect islandic motion, but the number of spectra in some occasions was smaller than the number of islands, because symmetric islands produces identical spectra. Using the $S(g)$ spectrum and particularly by adding the term $-r_ip_{ri}$ to its definition, we have managed to over through this drawback.

Figure 17a shows the $S(g)$ spectrum for a chaotic orbit. In this case, we see a complicated, asymmetric spectrum, with a lot of large and small peaks. This orbit belongs to the chaotic sea shown in Fig. 6. Initial conditions are: $r_0=3.4, p_{r0}=22, z_0=0$. The corresponding L.C.E for a time period of $10^5$ time units, is given in Fig. 17b.

In Fig. 16a, we have seen the $S(g)$ spectrum for an orbit producing the set of nine small small islands, shown in Fig. 4. Figure 18a shows the $S(g)$ spectrum for an orbit starting near the above described regular orbit. Initial conditions are: $r_0=9.605, p_{r0}=0, z_0=0$. Here, we can see again nine spectra each one corresponding to an island. The basic difference between Fig. 16a and 18a is that in Fig. 18a we observe some additional small peaks. Those additional peaks, indicate sticky motion. It is well known, that in a dynamical system of two degrees of freedom, sticky orbits are those orbits, which stay for long time periods near the last K.A.M torus before they escape to surrounding chaotic sea (see Karanis \& Caranicolas 2002).

Let us now start to follow the time evolution of a sticky orbit, using the new $S(g)$ spectrum. The sticky period is about 1750 time units. Figure 18b shows the $S(g)$ spectrum of the sticky orbit about 400 time units after the test particle has left the sticky region. Here the nine spectra have joined together producing three sets of spectra. Figure 18c shows the $S(g)$ spectrum of the sticky orbit about 700 time units after the test particle has left the sticky region. Here the nine spectra have joined all together producing a single spectrum. This spectrum has the characteristics of a chaotic spectrum and strongly indicates, that the test particle has left completely from the sticky region and has gone to the chaotic sea. The shape of the $S(g)$ spectrum after the test particle has traveled for 20000 time units in the chaotic sea, is shown in Figure 18d.

\section{A semi - theoretical approach}

In the present Section, we shall present some theoretical arguments, together with elementary numerical calculations, in order to explain the numerically found relationships, given in Figs. 9, 10 and 12. Since the potential of the nucleus (3) is integrable with spherical symmetry, the contribution of the total potential, to the chaotic regions observed in the $(r-p_r)$ phase planes of the dynamical system, should derive mainly from the $F_{zg}$ force, which is the vertical force in the $z$ - direction. The $F_{zg}$ force is
\begin{equation}
F_{zg}=-\frac{bz\left( a+\sqrt{{{h}^{2}}+b{{z}^{2}}} \right){{\upsilon }_{0}}^{2}}{\sqrt{{{h}^{2}}+b{{z}^{2}}}\left[ {{r}^{2}}+{{\left( a+\sqrt{{{h}^{2}}+b{{z}^{2}}} \right)}^{2}}+{{c}^{2}} \right]}.
\end{equation}

Figure 19 shows a plot of the $|F_{zg}|$ vs. $b$, for the elliptical and the disk galaxy models. As the scattering occurs near the nucleus, it must be at $r<1$ and $z<1$. The particular values of $r$ and $z$ are irrelevant. In this case, we choose: $r=r_0=0.2$ and $z=z_0=0.1$. We observe, that $|F_{zg}|$, increases almost linearly with $b$, in both cases. One can see that, the pattern between $|F_{zg}|$ and the flatness parameter $b$ (shown in Fig. 19), is similar to those between the maximal L.C.Es or the chaotic percentage $A\%$ and the flatness parameter $b$ (shown in Figs. 9 and 10 respectively), which have been obtained numerically. In both cases numerical outcomes suggest that, as the value of the flatness parameter $b$ increases, the chaoticity of the dynamical system increases as well. Numerically found relationships given in Figs. 9 and 10 can be explained, if $F_{zg}$ force is the main contributor, to the chaotic regions observed in the $(r-p_r)$ phase planes of the dynamical system.
\begin{figure}[!tH]
\centering
\resizebox{0.9\hsize}{!}{\rotatebox{0}{\includegraphics*{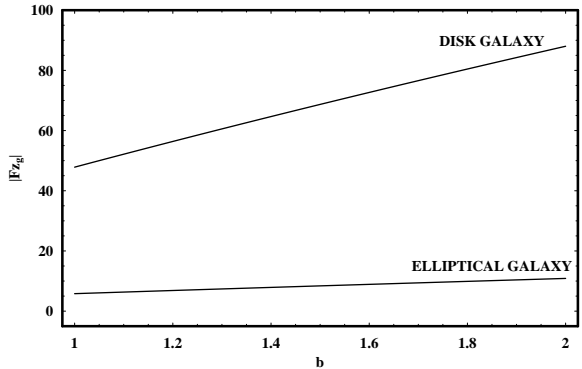}}}
\caption{A semi - theoretical plot of $|F_{zg}|$ vs. $b$, for the elliptical and disk galaxy models. The values of all other parameters are given in the text.}
\end{figure}

In order to explain the results shown in Fig. 12, we use an analysis similar to that used in Caranicolas \& Innanen (1991). When a star approaches the dense nucleus, its momentum in the $z$ - direction, changes according to the equation
\begin{equation}
m\Delta {{\upsilon }_{z}}=\left\langle {{F}_{zt}} \right\rangle \Delta t,
\end{equation}
where $m$ is the mass of the star, $\Delta t$ is the duration of the encounter and $<F_{zt}>$ is the total average $F_z$ force. It was observed, that stars deflection into higher $z$ proceeds in each case cumulatively, a little more with each successive pass by the nucleus and not with a single dramatic encounter. It is assumed, that the star is scattered off the galactic plane after $n$ $(n>1)$ encounters, when the total change in the momentum in the $z$ - direction, is of order of the tangential velocity $\upsilon _{\phi}=L_z/r$. Therefore we have
\begin{equation}
m\sum\limits_{i=0}^{n}{\Delta {{\upsilon }_{zi}}\approx \left\langle {{F}_{zt}} \right\rangle }\sum\limits_{i=0}^{n}{\Delta {{t}_{i}}}.
\end{equation}

Let
\begin{eqnarray}
m&=&1, \nonumber \\
\sum\limits_{i=0}^{n} \Delta \upsilon _{zi}&=&\frac{L_z}{r}, \nonumber \\
\sum\limits_{i=0}^{n} \Delta t_i&=&T_e, \nonumber \\
F_{zt} &=& F_{zn} + F_{zg} \ \ \ ,
\end{eqnarray}
where $F_{zn}$ stands for the constant vertical force, coming from the dense nucleus, computed at a given point and for fixed values of the involved parameters. Then Eq. (14) becomes
\begin{equation}
\frac{{{L}_z}}{r}\approx \left( F_{zn} + F_{zg} \right) {{T}_{e}}.
\end{equation}

Initially, if $\upsilon_z \geq \upsilon _ {\phi}$ the orbit is chaotic. Otherwise, $\upsilon_z \ll \upsilon _ {\phi}$ and after successive encounters with the galactic center, $\upsilon_z$ will increase, until it reaches $\upsilon_{\phi}$. Then the orbit will be chaotic. The time needed to reach the condition $\upsilon_z > \upsilon_{\phi}$ is $T > T_e$, where $T_e$ is the sum of the encounter durations (15). Let $T_f$ be the age of the galaxy, which corresponds here to the final time of numerical integrations. If $T_e > T_f$, then some stars do not have enough time to be scattered. We conclude from Eq. (16), that if $L_z > L_{zc}$, where $L_{zc}$ is the angular momentum computed with $T_e = T_f$, then there is still exist non-chaotic motions. Whereas for $L_z < L_{zc}$, all stars may have time to deploy chaotic motions. The star must go close to the nucleus, in order to be scattered. In this case, we may set $r=r_0<1$ and $z=z_0<1$ in (16) and then we conclude in
\begin{equation}
L_{zc}\thickapprox k + \lambda b,
\end{equation}
where
\begin{eqnarray}
k &=& F_{zn}T_f r_0, \nonumber \\
\lambda &=& \frac{F_{zg}T_f r_0}{b},
\end{eqnarray}
are constants, at a given point $P(r_0, z_0)$. Relation (17) is very simplistic, but nevertheless reproduces the correct dependency on the flattening parameter $b$. Specifically, this expression explains the linear relationship between the critical value of the angular momentum $L_{zc}$ and the flattening parameter $b$, shown in Fig. 12. The reader can find similar analysis in Caranicaolas \& Innanen (1991) and Karanis \& Caranicolas (2001). The different slope in Fig. 12, for the elliptical galaxy and disk galaxy model can be explained, because the values of $k$ and $\lambda$, of equation (17) are different for the two types of galaxies, depending on the remaining dynamical parameters, such as $\upsilon_0, a$ and $h$.

\section{Discussion and conclusions}

In the present paper, we have studied the properties of motion, in a new galactic dynamical model. This new model, is a combination of the logarithmic and the Miyamoto - Nagai model (1975). The choice seems to be very successful for several reasons. The first reason is that, by changing the appropriate values of the parameters , one can obtain a variety of galactic models. Another reason, is that the dynamical system is simpler and displays a large variety of families of orbits, compared to other models, such as the model proposed by Carlsberg and Innanen (1987), (see also Caranicolas \& Innanen 1991, Caranicolas 1997).

It is well known, that galaxies evolve. Today, two main methods are generally accepted, in order to explain the results obtained from observations. In the first method, galaxies evolve with little influence from their environment, while in the second, small galaxies assemble to form more complicated stellar systems (see Guiderdoni 2002 and references therein). Observations show that, elliptical galaxies are redder than disk galaxies, because they are made of old stars and there is no more gas in order to form young stars. On the contrary, disk galaxies are more blue, they contain younger stars and they have still gas to form them. Our galactic model, can be used in order to follow the galactic evolution from flat disk systems to elliptical ones. This can be obtained, using a time-dependent form of potential (1), in which the parameters $(\upsilon_0, a, h)$ varies with time. Indeed, it seems that the chaotic region is more significant in disk galaxies. As a result, stars can escape from the galactic disk, so as the mass distribution changes and looks like more and more that of an elliptical galaxy.

When studying Galactic Dynamics, it is invariably found that the stellar rotation velocity remains constant, or flat, with increasing distance away from the galactic center (see Catinella et al. 2006). This result, is highly counterintuitive since, based on Newton's Law of gravity, the rotational velocity would steadily decrease for stars further away from the galactic center. By this particular argument, the flat rotational curves, seem to imply that each galaxy must be surrounded by significant amounts of dark matter. It has been postulated and generally accepted, that the dark matter would have to be located in a massive, halo enshrouding each galaxy (see Caranicolas \& Zotos 2009, Caranicolas \& Zotos 2010). The first real surprise in the study of dark matter lay in the outermost parts of galaxies, known as galaxy halos. Here, there is negligible luminosity, yet there are occasional orbiting gas clouds, which allow one to measure rotation velocities and distances. The rotation velocity is found, not to decrease with increasing distance from the galactic center, implying that the mass distribution of the galaxy, cannot be considered, like the light distribution. The mass must continue to increase: since the rotation velocity satisfies $\Theta ^2=GM/r$, where $M$ is the mass within radius $r$, we infer that $M$ increases proportionally to $r$. This rise appears to stop at about $50 kpc$, where halos appear to be truncated. We infer that the mass - luminosity ratio of the galaxy, including its disk halo, is about 5 times larger than estimated for the luminous inner region, or equal to about 50.

The flat profile of the rotation curve shown in Fig. 1a, corresponding to our elliptical galaxy model, seems to be compatible with those of NGC 4261 (see Van der Marel, et al. 1990) and NGC 2778, NGC 3818, NGC 584, NGC 3557, NGC 1553, NGC 5866, NGC 4594 and NGC 3115 (see Davies et al. 1983). At the same time, the flat profile of the rotation curve shown in Fig. 1b, corresponding to our disk galaxy model, seems to be compatible with those of NGC 801, NGC 7331, NGC 224 (M31), NGC 2590, NGC 2841, UMa: NGC 3953, UMa: NGC 3992, NGC 5907, NGC 1097 and NGC 4321 (see Brownstein and Moffat 2006). Here we must clarify, that when we say that the flat profiles of the theoretical rotation curves seem to be compatible with those from real galaxies, we mean that, if we superpose Fig. 1a-b with the rotation curves from the references, we will see that the theoretical rotation curves of our present dynamical model pass through the observation data point, within the error bars. Additionally, both rotation curves shown in Figs. 1a-b, remain flat with increasing distance from the galactic center. In both cases, we see that the asymptotic behavior of the theoretical rotation curves is very similar to the referenced rotation curves obtained by observation data. In Fig. 1a-b we observe a bump for small values of $r$. This bump is caused, in both cases, by the dense spherical nucleus. Here we must note that, the above mentioned rotation curves of galaxies, were used in order to choose the appropriate values of the parameters $(\upsilon_0, a, h)$ in both the elliptical and disk galaxy model. The chosen values of these parameters, secure positive value of density everywhere. Therefore, our new dynamical model, is realistic and can be used in order to describe real elliptical and disk galaxies.

We have presented a description of the phase plane, for both the elliptical and the disk galaxy for low angular momentum stars, that is $L_z \leq 10$. The reason for doing that, is because in this case, we observe interesting families of regular orbits while, at the same time, a considerable part of the surface of section is covered by chaotic orbits. On the other hand, regular orbits have as conserved quantity some third integral, or quasi integral besides the energy and the $z$ - component of the angular momentum. The reader can find more detailed information about the behavior of low angular momentum stars in a series of papers (Caranicolas and Innanen 1991, Caranicolas 1997, Karanis \& Caranicolas 2001, Caranicolas \& Papadopoulos 2003, Caranicolas \& Zotos 2010). Here, we must emphasize that the phase planes shown in Figs. 4 and 6, show that a test particle (that is a star with a negligible mass), with energy $E=486$ in the elliptical galaxy model, shows the orbital behavior, displayed by the phase plane of Fig. 4, while a test particle with energy $E=1425$ in a disk galaxy, shows the orbital behavior displayed by the phase plane of Fig. 6. We have chosen these values of energy, in each case, in order to cover a phase plane of about the same distance from the center of the galaxy.

In order to keep things simple, we have kept some of the parameters of the system constant and studied the behavior of orbits varying only four basic parameters $\left(\upsilon_0, a, b, h \right)$. The parameter $\upsilon_0$ of our model, scales the rotation curve due to the potential $V_g$, while the triplet $\left(a, b, h \right)$ determines the density distribution and the structure of the dynamical system (elliptical or disk galaxy). It was found that, both in the elliptical and disk galaxy model, there is a linear relationship between the L.C.E and the flatness parameter $b$. On the contrary, the relation between the chaotic percentage $A\%$ of the phase plane and the flatness parameter $b$ is not linear but exponential. In both models, the numerically found outcomes were explained, using some semi - theoretical arguments. In the same sense, we have explained the numerically obtained relationship between the critical value of the angular momentum $L_{zc}$ and the flatness parameter $b$ of the dynamical system.

A very effective and reliable tool for the study of motion, is the new $S(g)$ spectrum, which allows us to detect islandic motion of resonant orbits, as it produces as much spectra as the total number of islands in the $(r-p_r)$ surface of section. One more advantage of this spectrum, is that it can be used in order to calculate the sticky period of an orbit and also to follow its time evolution towards the chaotic sea. In this point, we must point out that all previously used spectra cannot identify tiny islands of resonant orbits (see Fig. 16b). We have strong numerical evidence, that the new definition of the $S(g)$ spectrum has the ability to detect islandic motion of complicated resonant orbits, which produce multiple islands in the Poincar\'{e} surface of section. It is in our future plans, to use it in several and more complicated dynamical systems, with additional theoretical work, in order to check if it gives always the correct number of spectra regarding the number of islands in the Poincar\'{e} surface of section.

Finally, we would like to point out, that the chaotic phenomena, are more significant, in the case where a massive nucleus was present in the core of the galaxy and also the chaotic motion is observed only for the low angular momentum stars (see Caranicolas and Papadopoulos 2003). Moreover, our numerical investigation reveals that, the degree of chaos in disk galaxies is much larger than compared to that in elliptical galaxies. This result is of particular interest, due to the fact that is generally accepted today, that massive objects inhabit the core of most galaxies (see Sprurzem 2003).

\section*{Acknowledgments}

\textit{I would like to express my thanks to Professor N. D. Caranicolas for stimulating discussions, during this investigation. I also would like to thank the anonymous referee for the careful reading of this manuscript and also for his very useful suggestions and comments, which improved the quality of the present paper}.

\end{document}